\documentclass[review, times, 3p]{elsarticle}
	\biboptions{longnamesfirst, numbers}

\journal{Journal of Multivariate Analysis}

\usepackage{amssymb, amsmath, amsthm, amsbsy, bm}
	\allowdisplaybreaks[4]
	\newtheorem{theorem}{Theorem}
	
	\newtheorem{lemma}{Lemma}
	\newtheorem{proposition}{Proposition}
	\newtheorem{remark}{Remark}

\usepackage{lineno,hyperref}
\usepackage{algorithm, algpseudocode}
\usepackage{textcomp}
\usepackage{caption, subcaption}
\usepackage{enumitem}
	\setenumerate[1]{itemindent=\parindent, itemsep=0pt, partopsep=0pt, parsep=\parskip, topsep=5pt}
\usepackage{appendix}

\DeclareMathOperator*{\argmax}{arg\,max\,}
\DeclareMathOperator*{\argmin}{arg\,min\,}
\DeclareMathOperator*{\cov}{cov}
\DeclareMathOperator*{\dd}{d\!}
\DeclareMathOperator*{\DD}{D\!}
\DeclareMathOperator*{\E}{E}

\DeclareMathOperator*{\T}{T}
\DeclareMathOperator*{\var}{var}

\begin{document}

\begin{frontmatter}



\title{Functional Continuum Regression\tnoteref{mytitlenote}}
\tnotetext[mytitlenote]{
	Supported by the Natural Sciences and Engineering Research Council of Canada (NSERC).
}

\author[sfu]{Zhiyang Zhou\corref{mycorrespondingauthor}}
\ead[url]{http://www.sfu.ca/~zza115/}
\ead{zhiyang\_zhou@sfu.ca}


\cortext[mycorrespondingauthor]{Corresponding author}

\address[sfu]{Department of Statistics \& Actuarial Science, Simon Fraser University, BC V5A 1S6, Canada}

\begin{abstract}
	Functional principal component regression (PCR) can fail to provide good prediction
	if the response is highly correlated with some excluded functional principal component(s).
	This situation is common
	since the construction of functional principal components never involves the response.
	Aiming at this shortcoming, 
	we develop functional continuum regression (CR).
	The framework of functional CR includes,
	as special cases,
	both functional PCR and functional partial least squares (PLS).
	Functional CR is expected to own a better accuracy 
	than functional PCR and functional PLS
	both in estimation and prediction;
	evidence for this is provided through simulation and numerical case studies.
	Also,
	we demonstrate the consistency of estimators given by functional CR.

\end{abstract}


\end{frontmatter}


\section{Introduction}\label{sec:introduction}

\subsection{Scalar-on-function regression model}

With the development of technology,
the demand for functional data analysis (FDA) is increasing;
it is frequent to encounter data 
that are recorded continuously within a nondegenerate and compact interval ${\mathcal{T}}$.
Scalar-on-function regression models
link the scalar response $Y$ to 
the integral of product of 
random process $X=X(t)$, $t\in{\mathcal{T}}$,
and corresponding unknown fixed function $\beta=\beta(t)$.
To be precise,
the linkage is of the form that
\[
	Y=\E Y+\int_{\mathcal{T}}\beta(X-\E X)+\varepsilon,
\]
where 
$\int_{\mathcal{T}} f$
is short for the Lebesgue integral
$\int_{\mathcal{T}}f(t)\dd t$
and where
$\E(\varepsilon |X)=0$,
$\cov(\varepsilon, X)=0$,
and $\int_{\mathcal{T}}\E X^2=\E\left(\int_{\mathcal{T}} X^2 \right)<\infty$.
All the functions involved in this paper
are assumed to be square-integrable,
i.e., 
the discussion is limited to $L^2(\mathcal{T})$ (or $L^2(\mathcal{T}\times\mathcal{T})$),
the $L^2$-space on $\mathcal{T}$ (or $\mathcal{T}\times\mathcal{T}$) with respect to the Lebesgue measure.
In addition, 
$\|\cdot\|$ stands for the $L^2$-norm,
i.e.,
$\|f\|$ equals $\sqrt{\int_{\mathcal{T}}f^2}$ for $f\in L^2(\mathcal{T})$ 
and $\sqrt{\int_{\mathcal{T}}\int_{\mathcal{T}}f^2}$ if $f\in L^2(\mathcal{T}\times \mathcal{T})$.

The infinite-dimensional structure of functional space makes data analysis challenging:
the dimension of parameter space exceeds the number of observed subjects,
and hence dimension-reduction techniques are indispensable
in model-fitting.
To estimate the coefficient function $\beta$ and the conditional expectation
\[
	\eta(x)=\E(Y|X=X_0)=\E Y+\int_{\mathcal{T}}\beta(x-\E X)
\]
for $x$,
a realization of $X$,
the standard approach is to express $\beta$ in terms of a finite set of functions $\{w_1,\ldots,w_p\}$ 
truncated from basis functions $\{w_1, w_2,\ldots \}\subset L^2(\mathcal{T})$.
This inspires us to approximate $\beta$ and $\eta(x)$, respectively, by
\begin{equation}\label{eq:betap}
	\beta_p
	=\argmin_{\theta\in\text{span}\{w_1,\ldots,w_p\}}\E\left(Y-\E Y-\int_{\mathcal{T}}\theta(X-\E X)\right)^2
	=\argmin_{\theta\in\text{span}\{w_1,\ldots,w_p\}}\E\left(\int_{\mathcal{T}}(\beta-\theta)(X-\E X)\right)^2
\end{equation}
and
\begin{equation}\label{eq:hp}
	\eta_p(x)=\E Y+\int_{\mathcal{T}}\beta_p(x-\E X),
\end{equation}
where $\text{span}\{w_1,\ldots,w_p\}$ is the linear space spanned by $w_1,\ldots,w_p$.
Note that $\beta_p$
is the slope of the best approximation (restricted within $\text{span}\{w_1,\ldots,w_p\}$ and in the $L^2$-sense) 
to $Y$ by a linear function of $X$.
In particular,
$\beta_p=\sum_{j=1}^{p}\left(\int_{\mathcal{T}}\beta w_j\right) w_j$
if $\{w_1,\ldots,w_p\}$ is orthonormal 
(i.e., $\int_{\mathcal{T}}w_j^2=1$ for $1\leq j\leq p$ 
and $\int_{\mathcal{T}}w_j w_{j'}=0$ for $j\neq j'$).
In addition,
for the completeness of definition,
write $\beta_0=0$ and $\eta_0(x)=\E Y$.

Though it is possible to employ a basis independent of the data 
(e.g., polynomial basis, Fourier basis, etc.), 
it is more reasonable to force the basis adapt to data. 
In that case,
$w_1,,w_2,\ldots$ are often unknown apriori and 
need to be replaced with corresponding estimates $\hat{w}_1, \hat{w}_2,\ldots$. 
Let $(X_i,Y_i)$, $i=1,\ldots,n$, be $n$ pairs of independently observed data,
all distributed as $(X,Y)$.
Correspondingly, 
we have estimates for \eqref{eq:betap} and \eqref{eq:hp}, respectively,
\[
	\hat{\beta}_p
	=\argmin_{\theta\in\text{span}\{\hat{w}_1,\ldots,\hat{w}_p\}}
	\frac{1}{n}\sum_{i=1}^{n}\left(Y_i-\bar{Y}-\int_{\mathcal{T}}\theta\left(X_i-\bar{X}\right)\right)^2
\]
and
\begin{equation}\label{eq:hp.hat}
	\hat{\eta}_p(x)=\bar{Y}+\int_{\mathcal{T}}\hat{\beta}_p\left(x-\bar{X}\right),
\end{equation}
where $\bar{X}=n^{-1}\sum_{i=1}^{n}X_i$ and $\bar{Y}=n^{-1}\sum_{i=1}^{n}Y_i$.

Under this framework,
the accuracy of estimates $\hat{\beta}_p$ and $\hat{\eta}_p(x)$ 
varies with the choice of $\{\hat{w}_1,\hat{w}_2,\ldots\}$.
Two well-known options are discussed in \autoref{sec:fpc.fpls},
leading to functional principal component regression (PCR)
and functional partial least squares (PLS), 
respectively.

\subsection{Functional principal component and functional partial least squares bases}\label{sec:fpc.fpls}

Among all the dimension-reduction techniques,
functional PCR is the most prevailing one.
It is built on the functional principal component basis,
say $\{w_{1,\text{FPC}},w_{2,\text{FPC}},\ldots\}$,
constructed from covariance operator $V_X: L^2(\mathcal{T})\rightarrow L^2(\mathcal{T})$ given by
\[
	V_X(f)(\cdot)=\int_{\mathcal{T}}f(s)v_X(s,\cdot)\dd s,
	\quad\forall f\in L^2(\mathcal{T}),
\]
where $v_X(s,t)=\cov(X(s),X(t))$.
The assumption $v_X\in L^2(\mathcal{T}\times\mathcal{T})$
implies that $V_X$ is of Hilbert-Schmidt class
and hence possesses a countable number of eigenvalues, 
all real and nonnegative.
Specifically,
$w_{p,\text{FPC}}$ is taken as the $p$-th eigenfunction of $V_X$,
or equivalently,
given $w_{1,\text{FPC}},\ldots,w_{p-1,\text{FPC}}$,
\begin{equation}\label{eq:wp.fpc}
	w_{p,\text{FPC}}
	=\argmax_w\int_{\mathcal{T}}wV_X(w)
\end{equation}
subject to
\[
	\|w\|=1
	\quad\text{and}\quad
	\int_{\mathcal{T}}ww_{j,\text{FPC}}=0,
	\quad 1\leq j<p.
\]
Empirically,
substitute 
\[
	\hat{v}_X(s,t)
	=\widehat{\cov}(X(s),X(t))
	=\frac{1}{n}\sum_{i=1}^n\left(X_i(s)-\bar{X}(s)\right)\left(X_i(t)-\bar{X}(t)\right)
\]
for $v_X(s,t)$ and 
then estimate $w_{p,\text{FPC}}$ by $\hat{w}_{p,\text{FPC}}$,
the $p$-th eigenfunction of operator $\widehat{V}_X$ satisfying that
\begin{equation}\label{eq:v.hat}
	\widehat{V}_X(f)(\cdot)=\int_{\mathcal{T}}f(s)\hat{v}_X(s,\cdot)\dd s,
	\quad\forall f\in L^2(\mathcal{T}).
\end{equation}

During the past few decades, 
extensive work has focused on functional PCR;
more details can be found in a number of monographs (e.g., \citet{RamsaySilverman2005} and \citet{HorvathKokoszka2012})
and review papers (e.g., \citet{WangChiouMuller2016} and \citet{Febrero-BandeGaleanoGonzalez-Manteiga2017}).

As defined in \eqref{eq:wp.fpc},
the construction of the functional principal component basis is ``unsupervised'' due to no involvement of $Y$;
the first $p_0$ elements of this basis seek to explain most of the variation of $X$,
whereas they are not necessarily important in representing $\beta$.
That is,
it is possible for one or more elements in the abandoned part 
$\{w_{p+1,\text{FPC}},\ldots\}$
to be highly correlated with the response.

Some efforts have already been made to target this well-known defect,
including \citet{PredaSaporta2005} who extended the multivariate PLS to functional PLS.
This technique relies on functional PLS basis which is defined in a sequential manner:
given $w_{1,\text{FPLS}},\ldots,w_{p-1,\text{FPLS}}$,
\[
	w_{p,\text{FPLS}}=\argmax_w{\cov}^2\left(Y-\eta_{{p-1},\text{FPLS}}(X),\int_{\mathcal{T}}Xw\right)
\]
subject to
\[
	\|w\|=1
	\quad\text{and}\quad
	\int_{\mathcal{T}}wV_X(w_{j,\text{FPLS}})=0,
	\quad 1\leq j<p.
\]
The corresponding empirical version is
\[
	\hat{w}_{p,\text{FPLS}}
	=\argmax_w
		\left(\frac{1}{n}\sum_{i=1}^n\left(Y_i-\hat{\eta}_{{p-1},\text{FPLS}}(X_i)\right)
		\left(\int_{\mathcal{T}}\left(X_i-\bar{X}\right)w\right)\right)^2
\]
subject to
\[
	\|w\|=1
	\quad\text{and}\quad
	\int_{\mathcal{T}}w\widehat{V}_X(\hat{w}_{j,\text{FPLS}})=0,
	\quad 1\leq j<p,
\]
where functions $\eta_{{p-1},\text{FPLS}}$ and $\hat{\eta}_{{p-1},\text{FPLS}}$ 
are respective counterparts of \eqref{eq:hp} and \eqref{eq:hp.hat}.

Functional PLS has been later investigated and developed by,
for instance,
\citet{ReissOgden2007, DelaigleHall2012, AguileraAguilera-MorilloPreda2016}.
PLS and its derivatives are referred to as ``fully supervised'' and may suffer the ``double-dipping'' problem:
they employ the covariance between $Y$ and $X$ both for the construction of basis functions and for further prediction;
the resulting findings are possibly vulnerable and sensitive to small signals;
see \cite{Jung2018}.
\citet{NieWangLiuCao2018} attempted to put forward a linear combination of functional PCR and functional PLS;
their proposal lies between unsupervised and fully supervised techniques.
Different from these authors,
we borrow the idea of continuum regression (CR) from \citet{StoneBrooks1990}
and extend it to the learning of functional data.

\subsection{Continuum regression}

Briefly,
in the context of multivariate analysis
with response $\bm{y}\in\mathbb{R}^{n\times 1}$
and design matrix $\mathbf{X}\in\mathbb{R}^{n\times d}$
both column-mean-centered,
CR projects $\bm{y}$ to the linear space spanned by mutually orthogonal regressors 
$\mathbf{X}\bm{w}_{1,\alpha},\ldots,\mathbf{X}\bm{w}_{p,\alpha}\in\mathbb{R}^{n\times1}$, 
after successively computing
\begin{equation}\label{eq:wp.c}
	\bm{w}_{j,\alpha}
		=\underset{\substack{\bm{w}\in\mathbb{R}^{d\times 1}, \bm{w}^{\T}\bm{w}=1 \\ 
				\bm{w}_{j,\alpha}^{\T}\mathbf{X}^{\T}\mathbf{X}\bm{w}=0, \forall j'<j}}{\argmax}
			\left(\bm{w}^{\T}\mathbf{X}^{\T}\bm{y}\right)^2
			\left(\bm{w}^{\T}\mathbf{X}^{\T}\mathbf{X}\bm{w}\right)^{\frac{\alpha}{1-\alpha}-1},
\end{equation}
where $\alpha\in [0,1)$ and $p$ ($\leq d$) are to be tuned.
The most appealing property of CR,
as proved by \citet{StoneBrooks1990},
is that its framework encompasses ordinary least square (OLS) ($\alpha=0$),
PLS ($\alpha=1/2$),
and PCR ($\alpha\rightarrow 1$).
Accordingly,
the model resulting from CR is expected to outperform those from OLS, PLS, and PCR in terms of prediction.

There have been some further developments of CR.
\citet{Sundberg1993} connected it to the ridge regression.
\citet{BjorkstromSundberg1999} and \citet{Jung2018} revealed the analytical form of \eqref{eq:wp.c}.
\citet{LeeLiu2013} combined CR with the kernel learning to accommodate the nonlinear regression.
\citet{ChenCook2010} proved the possible inconsistency of estimators produced by CR,
while \citet{ChenZhu2015} showed the consistency given by CR 
in estimating the central (dimensional-reduction) subspace defined by \citet{Cook1996} and \citet[pp. 105]{Cook1998}. 

In the remainder of this paper, 
\autoref{sec:fcr} introduces functional CR and its special cases.
Our consistency results are presented in \autoref{sec:theory},
based on which \autoref{sec:implement} derives an effective algorithm.
Empirical evidence appears in \autoref{sec:numerical},
where our method is compared with existing ones in terms of both estimation and prediction.
\autoref{sec:conclusion} discusses of pros and cons of functional CR as well as possible future work.
For the sake of brevity,
technical details are left in appendices.

\section{Functional continuum regression}\label{sec:fcr}

\subsection{Functional continuum basis}

We begin by defining the functional continuum basis
which will be denoted by $\{w_{1,\alpha},w_{2,\alpha},\ldots\}$.
For a pre-determined $\alpha\in[0,1)$,
we construct the basis in a sequential way.
Given $w_{1,\alpha},\ldots,w_{p-1,\alpha}$,
define
\begin{equation}\label{eq:wp.fc}
	w_{p,\alpha}=\argmax_wT_{\alpha}(w)
\end{equation}
subject to 
\begin{equation}\label{eq:fcr.constraint}
	\|w\|=1
	\quad\text{and}\quad
	\int_{\mathcal{T}}wV_X(w_{j,\alpha})=0,
	\quad 1\leq j<p,
\end{equation}
where
\begin{equation}\label{eq:T}
	T_{\alpha}=T_{\alpha}(w)
	={\cov}^2\left(Y,\int_{\mathcal{T}}Xw\right)\cdot
	\left(\int_{\mathcal{T}}wV_X(w)\right)^{\frac{\alpha}{1-\alpha}-1}.
\end{equation}
Following \eqref{eq:betap} and \eqref{eq:hp}, 
\begin{align*}
	\notag\beta_{p,\alpha}
	&=\argmin_{\theta\in\text{span}\{w_{1,\alpha},\ldots,w_{p,\alpha}\}}
	\E\left(\int_{\mathcal{T}}(\beta-\theta)(X-\E X)\right)^2 \\
	&=\sum_{j=1}^p
		\left(\int_{\mathcal{T}}\beta V_X(w_{j,\alpha})\right)
		\left( \int_{\mathcal{T}}w_{j,\alpha}V_X(w_{j,\alpha}) \right)^{-1/2}
		w_{j,\alpha}
\end{align*}
and
\begin{align*}
	\eta_{p,\alpha}(x)
	&=\E Y+\int_{\mathcal{T}}\beta_{p,\alpha}(x-\E X) \\
	&=\E Y+
	\sum_{j=1}^p
		\left(\int_{\mathcal{T}}\beta V_X(w_{j,\alpha})\right)
		\left(\int_{\mathcal{T}}(x-\E X)w_{j,\alpha}\right)
		\left( \int_{\mathcal{T}}w_{j,\alpha}V_X(w_{j,\alpha}) \right)^{-1/2}
\end{align*}
are resulting approximations to $\beta$ and $\eta(x)$, respectively. 

Having defined the population version of the functional continuum basis functions,
we now give the empirical counterpart.
The empirical version, 
say $\{\hat{w}_{1,\alpha},\hat{w}_{2,\alpha},\ldots\}$,
is defined recursively.
Once the first $j-1$ elements are determined,
$\hat{w}_{p,\alpha}$ is taken as 
the maximizer of following optimization problem:
\begin{equation}\label{eq:T.hat}
	\begin{aligned}
		&\underset{w}{\text{maximize}} & & 
			\widehat{T}_{\alpha}(w)=
			\left(\frac{1}{n}\sum_{i=1}^n(Y_i-\bar{Y})\left(\int_{\mathcal{T}}(X_i-\bar{X})w\right)\right)^2
			\left(\int_{\mathcal{T}}w\widehat{V}_X(w)\right)^{\frac{\alpha}{1-\alpha}-1} \\
		&\text{subject to} & & \|w\|=1
			\quad\text{and}\quad
			\int_{\mathcal{T}}w\widehat{V}_X(\hat{w}_{j,\alpha})=0 
			\quad 1\leq j<p, 
	\end{aligned}
\end{equation}
where operator $\widehat{V}_X$ is defined as \eqref{eq:v.hat}.
Further,
$\beta_{p,\alpha}$ and $\eta_{p,\alpha}(x)$ are respectively estimated by
\begin{align}
	\hat{\beta}_{p,\alpha}
	&=\argmin_{\theta\in\text{span}\{\hat{w}_{1,\alpha},\ldots,\hat{w}_{p,\alpha}\}}
	\frac{1}{n}\sum_{i=1}^{n}\left(Y_i-\bar{Y}-\int_{\mathcal{T}}\left(X_i-\bar{X}\right)\theta\right)^2 \notag\\
	&=\sum_{j=1}^p
		\left(\int_{\mathcal{T}}\beta \widehat{V}_X(\hat{w}_{j,\alpha})\right)
		\left( \int_{\mathcal{T}}\hat{w}_{j,\alpha}\widehat{V}_X(\hat{w}_{j,\alpha}) \right)^{-1/2}
		\hat{w}_{j,\alpha} \notag\\
	&=\sum_{j=1}^p
		\widehat{\cov}\left( Y, \int_{\mathcal{T}}X\hat{w}_{j,\alpha} \right)
		\widehat{\var}^{-\frac{1}{2}}\left(
			\int_{\mathcal{T}}X\hat{w}_{j,\alpha}
		\right)
		\hat{w}_{j,\alpha} \label{eq:beta.hat.fc}
\end{align}
and
\begin{equation}\label{eq:h.hat.fc}
	\hat{\eta}_{p,\alpha}(X)=
	\bar{Y}+\int_{\mathcal{T}}\hat{\beta}_{p,\alpha}\left(x-\bar{X}\right).
\end{equation}

Return to the definition of $w_{p,\alpha}$ in \eqref{eq:wp.fc}.
Though it looks like a natural extension of that of the $d$-vector \eqref{eq:wp.c},
at least two concerns
(Propositions \ref{prop:existence} and \ref{prop:expand})
arise with the non-concavity of 
objective functions $T_{\alpha}(w)$ and $\widehat{T}_{\alpha}(w)$
and the infinite dimension of $L^2(\mathcal{T})$:
one is the existence of $w_{p,\alpha}$ and $\hat{w}_{p,\alpha}$
which is not trivial at all
since the unit sphere and unit ball in $L^2(\mathcal{T})$ are no longer compact;
the other is whether or not,
for any preset $\alpha\in[0,1)$,
$\beta$ can be fully expressed in terms of 
the functional continuum basis $\{w_{1,\alpha},w_{2,\alpha},\ldots\}$.

\begin{proposition}\label{prop:existence}
	Given $w_{1,\alpha},\ldots,w_{p-1,\alpha}$,
	the objective function $T_{\alpha}$ defined in \eqref{eq:T}, 
	subject to conditions \eqref{eq:fcr.constraint},
	has a maximizer. 
	So does $\widehat{T}_{\alpha}$ in \eqref{eq:T.hat} with fixed $\hat{w}_{1,\alpha},\ldots,\hat{w}_{p-1,\alpha}$.
\end{proposition}

\begin{proposition}\label{prop:expand}
	Suppose $\beta$ can be expansed in terms of eigenfunctions of $V_X$.
	Then, for each $\alpha\in[0,1)$,
	$\beta$ belongs to $\overline{\text{span}\{w_{1,\alpha},w_{2,\alpha},\ldots\}}$,
	the closure of $\text{span}\{w_{1,\alpha},w_{2,\alpha},\ldots\}$.
	 
\end{proposition}

\subsection{Special cases}\label{sec:special}

Functional CR inherits the inclusion property of CR;
i.e.,
for certain $\alpha$'s,
functional CR reduces to some existing methods.

Firstly, 
as $\alpha\rightarrow 1$,
the variance term $\int_{\mathcal{T}}wV_X(w)$
dominates the objective function \eqref{eq:T}
and the role of $\cov(Y, \int_{\mathcal{T}} Xw)$ is negligible.
We assert that,
in this scenario,
the functional continuum basis is identical to the functional principal component basis.

\begin{proposition}\label{prop:alpha->1}
	If $\cov\left(Y,\int_{\mathcal{T}}Xw_{p,\text{FPC}}\right)\neq 0$ for all $p\in\mathbb{N}$,
	then $w_{p,\alpha}=w_{p,\text{FPC}}$ as $\alpha\rightarrow 1$.
\end{proposition}

At the other extreme ($\alpha=0$),
note that
\[
	w_{1,0}
	=\underset{w:\|w\|=1}{\argmax}
	\frac{{\cov}^2\left(Y,\int_{\mathcal{T}}Xw\right)}{\int_{\mathcal{T}}wV_X(w)}
	=\underset{w:\|w\|=1}{\argmax}
	\frac{{\cov}^2\left(Y,\int_{\mathcal{T}}Xw\right)}{\var(Y)\var\left(\int_{\mathcal{T}} Xw\right)}.
\]
Geometrically, 
$w_{1,\alpha}$ maximizes the squared cosine of angle between 
$\int_{\mathcal{T}} Xw$ and $Y$.
Therefore,
$\int_{\mathcal{T}} Xw_{1,0}$ is parallel to 
the orthogonal projection of $Y$ onto $X$,
meaning that $\cov\left(Y,\int_{\mathcal{T}}Xw\right)$ must be zero
for all $w$ such that 
$\int_{\mathcal{T}} wV_X(w_{1,0})=0$.
That is to say,
the sequential construction terminates 
at $w_{1,\text{FCR}}$
and no subsequent element exists.
Obviously,
in this situation,
functional CR is equivalent to 
a functional version of OLS regression.

Another special case lies midway between two extremes, 
i.e., $\alpha=1/2$.
Noticing that under the need of constraints \eqref{eq:fcr.constraint},
we have
\[
	\cov\left(Y-\eta_{p-1,\frac{1}{2}}(X), \int_{\mathcal{T}}Xw\right)=\cov\left(Y, \int_{\mathcal{T}}Xw\right).
\]
One can see that 
this case is identical to the functional PLS introduced in \autoref{sec:fpc.fpls}.

\section{Theoretical properties}\label{sec:theory}

\subsection{Equivalent forms of the functional continuum basis}

Considering residuals of $X$ and $Y$ after the first $p-1$ steps,
we merge the last $p-1$ side-conditions in \eqref{eq:fcr.constraint}
and the objective function \eqref{eq:wp.fc} together.
This reformulation simplifies forthcoming proofs
and facilitates the implementation in \autoref{sec:implement} as well. 

\begin{proposition}\label{prop:equivalent.w}
	Given $w_{1,\alpha},\ldots,w_{p-1,\alpha}$ with 
	$\int_{\mathcal{T}}w_{j,\alpha}V_X(w_{j,\alpha})>0$ for $j=1,\ldots,p-1$,
	write 
	\[
		X^{(p,\alpha)}=X-\E X-
			\sum_{j=1}^{p-1}
				\left( \int_{\mathcal{T}}(X-\E X)w_{j,\alpha} \right)
				\left( \int_{\mathcal{T}}w_{j,\alpha}V_X(w_{j,\alpha}) \right)^{-1/2}
				V_X(w_{j,\alpha})
	\]
	and
	\[
		Y^{(p,\alpha)}=Y-\eta_{p-1,\alpha}(X)=\int_{\mathcal{T}}\beta X^{(p,\alpha)}.
	\]
	Then, $w_{p,\alpha}$ defined in \eqref{eq:wp.fc} can be found by maximizing $T_{p,\alpha}^*$ on the unit sphere,
	i.e.,
	\begin{equation}\label{eq:equivalent.w.hat}
		w_{p,\alpha}=\argmax_{w:\|w\|=1} T_{p,\alpha}^*(w),
	\end{equation}
	where
	\begin{align*}
		T_{p,\alpha}^*(w)
			&={\cov}^2\left(Y^{(p,\alpha)}, \int_{\mathcal{T}}X^{(p,\alpha)}w\right)\cdot
				\left(
					\int_{\mathcal{T}}wV_{X^{(p,\alpha)}}(w)
				\right)^{\frac{\alpha}{1-\alpha}-1}\\
			&=\left(\int_{\mathcal{T}}\beta V_{X^{(p,\alpha)}}(w)\right)^2
				\left(
					\int_{\mathcal{T}}wV_{X^{(p,\alpha)}}(w)
				\right)^{\frac{\alpha}{1-\alpha}-1}.
	\end{align*}
\end{proposition}

An empirical counterpart of \autoref{prop:equivalent.w} naturally follows.

\begin{proposition}\label{prop:equivalent.w.hat}
	Given $\hat{w}_{1,\alpha},\ldots,\hat{w}_{p-1,\alpha}$ with 
	$\int_{\mathcal{T}}\hat{w}_{j,\alpha}\widehat{V}_X(\hat{w}_{j,\alpha})>0$ for all $j\leq p-1$,
	write 
	\[
		\widehat{X}_i^{(p,\alpha)}
		=X_i-\bar{X}-
			\sum_{j=1}^{p-1}
				\left(\int_{\mathcal{T}}\left(X_i-\bar{X}\right)\hat{w}_{j,\alpha}\right)
				\left( \int_{\mathcal{T}}\hat{w}_{j,\alpha}\widehat{V}_X(\hat{w}_{j,\alpha}) \right)^{-1/2}
				\widehat{V}_X(\hat{w}_{j,\alpha})
	\]
	and
	\[
		\widehat{Y}_i^{(p,\alpha)}
		=Y_i-\hat{\eta}_{p-1,\alpha}(X_i)
		=\int_{\mathcal{T}}\beta\widehat{X}_i^{(p,\alpha)},
	\]
	for $i=1,\ldots,n$. 
	Then,
	\begin{equation}\label{eq:equivalent}
		\hat{w}_{p,\alpha}=\argmax_{w:\|w\|=1}\widehat{T}_{p,\alpha}^*(w),
	\end{equation}
	where
	\begin{align}
		\widehat{T}_{p,\alpha}^*(w)
			&=\widehat{\cov}^2\left(\widehat{Y}^{(p,\alpha)}, \int_{\mathcal{T}}\widehat{X}^{(p,\alpha)}w\right)\cdot
				\left(
					\int_{\mathcal{T}}w\widehat{V}_{\widehat{X}^{(p,\alpha)}}(w)
				\right)^{\frac{\alpha}{1-\alpha}-1} \notag \\
			&=\left(\int_{\mathcal{T}}\beta\widehat{V}_{\widehat{X}^{(p,\alpha)}}(w)\right)^2
				\left(
					\int_{\mathcal{T}}w\widehat{V}_{\widehat{X}^{(p,\alpha)}}(w)
				\right)^{\frac{\alpha}{1-\alpha}-1} \label{eq:T.p.star.hat}
	\end{align}
	with 
	$\widehat{V}_{\widehat{X}^{(p,\alpha)}}
	=\widehat{V}_{\widehat{X}^{(p,\alpha)}}(s,t)
	=n^{-1}\sum_{i=1}^n\widehat{X}_i^{(p,\alpha)}(s)\widehat{X}_i^{(p,\alpha)}(t)$.
\end{proposition}

Previously, 
both in \eqref{eq:wp.fc} and \eqref{eq:equivalent},
the functional continuum basis has been defined as a set of maximizers of sequential optimization problems.
\autoref{prop:explicit.w} derives an alternative but more explicit form of these desired solutions:
they are constructed by adjusting the projection of function $\beta$ on some directions.

\begin{proposition}\label{prop:explicit.w}
	Given $\alpha\in[0,1)$, $p\in\mathbb{N}$,
	and $w_{1,\alpha},\ldots,w_{p-1,\alpha}$.
	Let $\lambda_j^{(p,\alpha)}$ denote the $j$-th largest eigenvalue of $V_{X^{(p,\alpha)}}$ 
	with corresponding eigenfunction $\phi_j^{(p,\alpha)}$.
	Suppose $\lambda_1^{(p,\alpha)}$ has multiplicity $m\geq 1$,
	i.e., $\lambda_1^{(p,\alpha)}=\cdots=\lambda_m^{(p,\alpha)}>\lambda_{m+1}^{(p,\alpha)}$.
	If $V_{X^{(p,\alpha)}}(\beta)$ is not orthogonal to 
	$\text{span}\left\{\phi_1^{(p,\alpha)},\ldots,\phi_m^{(p,\alpha)}\right\}$,
	there exists $\delta^{(p,\alpha)}\in(-1,0)\cup(0,\infty)$
	such that
	\[
		w_{p,\alpha}\propto
		\sum_{j=1}^{\infty}
			\frac{\lambda_j^{(p,\alpha)}\left(\int_{\mathcal{T}}\beta\phi_j^{(p,\alpha)}\right)}
				{\lambda_j^{(p,\alpha)}+\lambda_1^{(p,\alpha)}/\delta^{(p,\alpha)}}
			\phi_j^{(p,\alpha)},
	\]
	where .
	The three boundary values of $\delta^{(p,\alpha)}$,
	$\{-1,0,\infty\}$,
	correspond to functional PCR ($\delta^{(p,\alpha)}\rightarrow -1$), 
	functional PLS ($\delta^{(p,\alpha)}\rightarrow 0$)
	and functional OLS ($\delta^{(p,\alpha)}\rightarrow\infty$),
	respectively.
\end{proposition}

\subsection{Consistency of the empirical functional continuum basis and corresponding estimators}

We need two more conditions:
\begin{enumerate}[label=(C\arabic*)]
	\item\label{cond:unique} For each $j$ ($\leq p$),
		$T_{j,\alpha}^*(w)$ attains a unique maximizer (up to sign) in $\{w\in L^2(\mathcal{T}):\|w\|=1\}$. 
	\item\label{cond:finite.moment}
		$\E\|X\|^4<\infty$.
\end{enumerate}
Our main result, \autoref{thm:consistence},
demonstrates the consistency of estimators in the case of ``fixed $p$ and infinite $n$''.

\begin{theorem}\label{thm:consistence}
	Fix $\alpha$ and $p$.
	Under \ref{cond:unique},
	$\|\hat{w}_{p,\alpha}-w_{p,\alpha}\|\stackrel{\text{P}}{\longrightarrow}0$ as $n\rightarrow\infty$.
	If we also have \ref{cond:finite.moment} apart from \ref{cond:unique},
	then 
	$\|\hat{\beta}_{p,\alpha}-\beta_{p,\alpha}\|$
	and $|\hat{\eta}_{p,\alpha}(x)-\eta_{p,\alpha}(x)|$
	both converge to zero in probability as $n$ diverges,
	where $x$ is a realization of $X$ and independent from $X_1,\ldots,X_n$.
\end{theorem}

\begin{remark}
	We do not have to impose uniqueness on the empirical version
	$\argmax_{\|w\|=1} \widehat{T}_{j,\alpha}^*(w)$ for all $j\leq p$.
	If $\argmax_{\|w\|=1} \widehat{T}_{j,\alpha}^*(w)$ is not unique,
	the proof of \autoref{thm:consistence} is still valid 
	as long as the chosen $\hat{w}_{j,\alpha}$ is measurable.
	\citet[Lemma 2]{Jennrich1969} provided a way to find such a measurable $\hat{w}_{j,\alpha}$.
\end{remark}

\section{Implementation}\label{sec:implement}

We understand that,
when facing real data,
it is not feasible to calculate the integrals exactly;
instead,
one may replace all integrals with discrete summations
after expanding $X_1,\ldots,X_n$ in terms of finite basis functions (e.g., B-splines).
Though this approximation definitely affects the accuracy of resulting estimators,
the corresponding discussion is out of the scope of this paper.
Therefore,
we will keep the integral notations, 
even in the description of implementation.

It is feasible to duplicate the ideas of \citet{LeeLiu2013} and \citet{ChanMak1990} 
to solve maximization problem \eqref{eq:T.hat} through a univariate root-finding procedure.
Nevertheless,
the implementation would be more natural and straightforward 
if we apply the following identity,
an empirical version of \autoref{prop:explicit.w}.

\begin{proposition}\label{prop:explicit.w.hat}
	Fix $\hat{w}_{1,\alpha},\ldots,\hat{w}_{p-1,\alpha}$.
	Let $\hat{\lambda}_j^{(p,\alpha)}$ is the $j$-th largest eigenvalue of 
	$\widehat{V}_{\widehat{X}^{(p,\alpha)}}$ with corresponding eigenfunction $\hat{\phi}_j^{(p,\alpha)}$.
	Suppose $\hat{\lambda}_1^{(p,\alpha)}=\cdots=\hat{\lambda}_m^{(p,\alpha)}>\hat{\lambda}_{m+1}^{(p,\alpha)}$.
	If 
	$\widehat{V}_{\widehat{X}^{(p,\alpha)}}(\beta)
	=n^{-1}\sum_{i=1}^n\widehat{X}_i^{(p,\alpha)}\widehat{Y}_i^{(p,\alpha)}$
	is not orthogonal to $\text{span}\left\{\hat{\phi}_1^{(p,\alpha)},\ldots,\hat{\phi}_m^{(p,\alpha)}\right\}$,
	there is $\hat{\delta}^{(p,\alpha)}\in(-1,0)\cup(0,\infty)$
	such that
	\begin{equation}\label{eq:explicit.w.hat}
		\begin{aligned}
			\hat{w}_{p,\alpha}=
			\left(
				\sum_{j=1}^{\infty}
				\frac{\widehat{\cov}^2\left(
						\widehat{Y}^{(p,\alpha)},
						\int_{\mathcal{T}}\widehat{X}^{(p,\alpha)}\hat{\phi}_j^{(p,\alpha)}
					\right)}
					{\left(\hat{\lambda}_j^{(p,\alpha)}+\hat{\lambda}_1^{(p,\alpha)}\left/\hat{\delta}^{(p,\alpha)}\right.\right)^2}
			\right)^{-\frac{1}{2}} 
			\sum_{j=1}^{\infty}
			\frac{\widehat{\cov}\left(
					\widehat{Y}^{(p,\alpha)},
					\int_{\mathcal{T}}\widehat{X}^{(p,\alpha)}\hat{\phi}_j^{(p,\alpha)}
				\right)}
				{\hat{\lambda}_j^{(p,\alpha)}+\hat{\lambda}_1^{(p,\alpha)}\left/\hat{\delta}^{(p,\alpha)}\right.}
			\hat{\phi}_j^{(p,\alpha)}.
		\end{aligned}
	\end{equation}	
\end{proposition}

\begin{remark}
	It is indispensable to assume that 
	$\widehat{V}_{\widehat{X}^{(p,\alpha)}}(\beta)$ 
	is not orthogonal to $\text{span}\left\{\hat{\phi}_1^{(p,\alpha)},\ldots,\hat{\phi}_m^{(p,\alpha)}\right\}$
	when $\hat{\lambda}_1^{(p,\alpha)}=\cdots=\hat{\lambda}_m^{(p,\alpha)}>\hat{\lambda}_{m+1}^{(p,\alpha)}$.
	So \autoref{prop:explicit.w.hat} does not cover all the possibilities;
	the ridge type solution may be not a global maximizer when the assumption is violated.
	Exceptions are constructible artificially,
	yet they are rare in practice \citep{BjorkstromSundberg1999,Jung2018},
	especially when $\varepsilon$ and $X(t)$ with given $t$ are all continuously distributed.
	Actually,
	if the assumption is not fulfilled,
	one can always project $\widehat{X}_i^{(p,\alpha)}$ 
	onto the (orthogonal) compliment of $\text{span}\left\{\hat{\phi}_1^{(p,\alpha)},\ldots,\hat{\phi}_m^{(p,\alpha)}\right\}$
	and take the projection as the substitute for $\widehat{X}_i^{(p,\alpha)}$.
\end{remark}

\begin{figure}
	\centering
	\begin{subfigure}{.5\textwidth}
		\centering
		\includegraphics[width=\textwidth, height=.4\textheight]{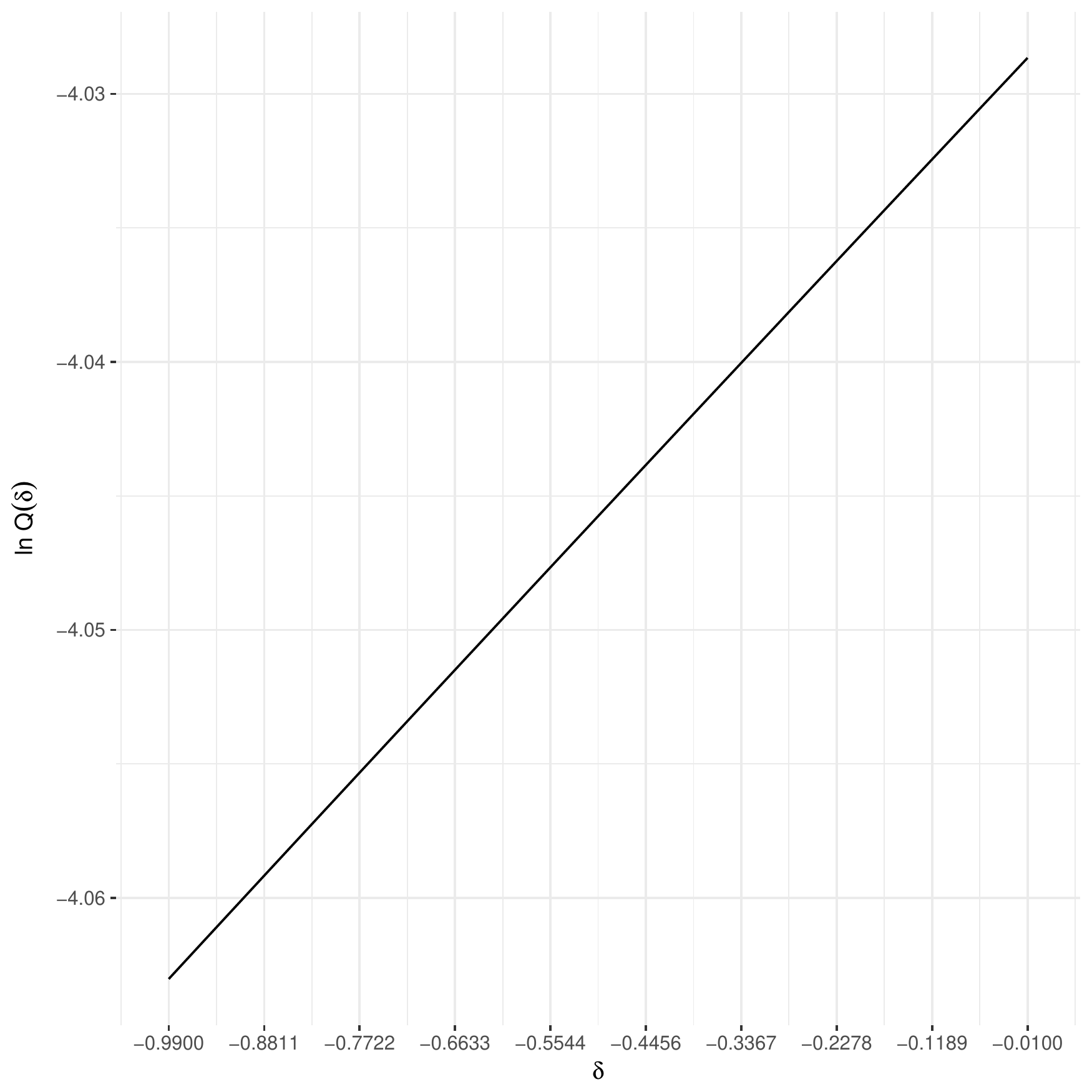}
		\caption{The left half ($-1<\delta<0$) of $\ln Q_{2,0}$}
	\end{subfigure}%
	\begin{subfigure}{.5\textwidth}
		\centering
		\includegraphics[width=\textwidth, height=.4\textheight]{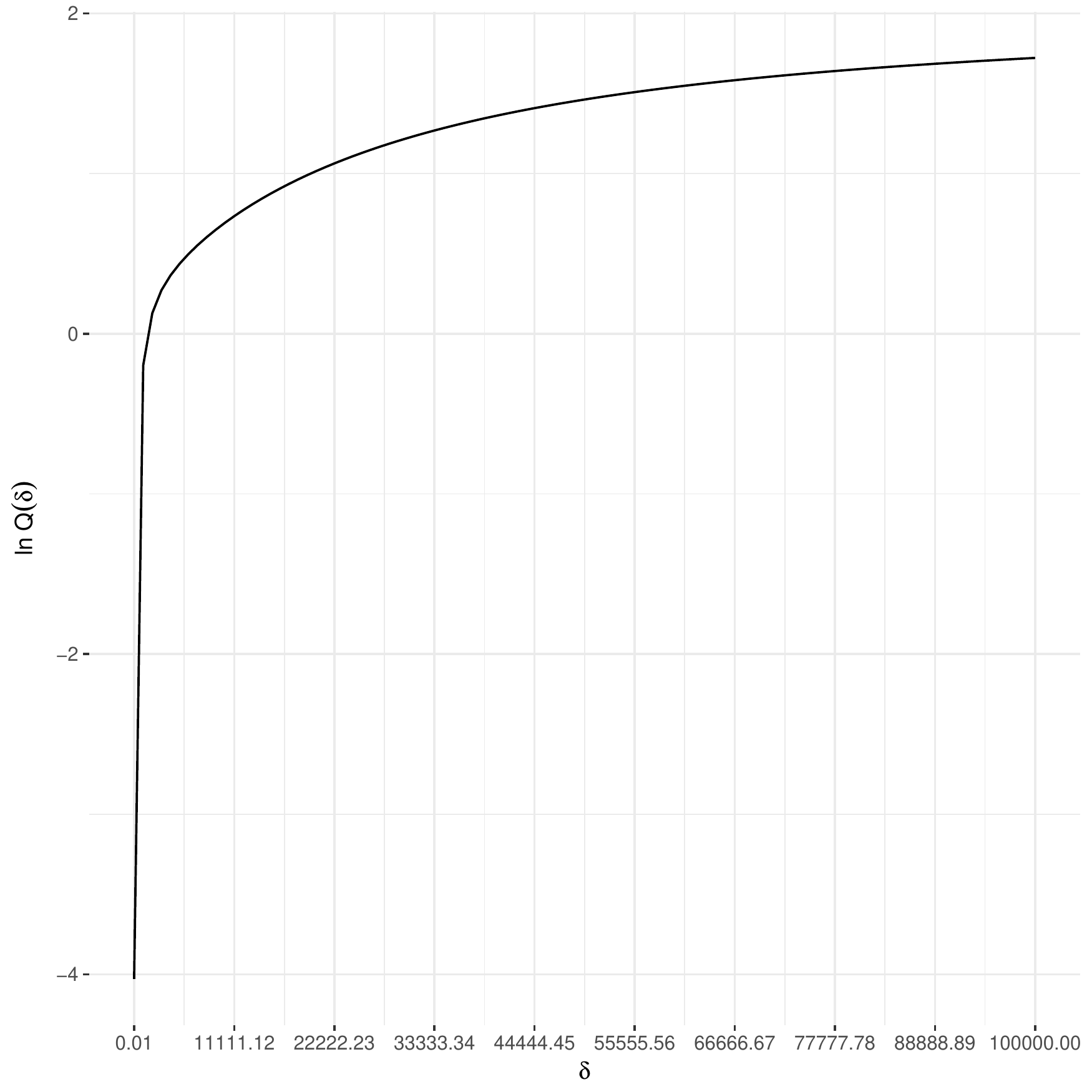}
		\caption{The right half ($\delta>0$) of $\ln Q_{2,0}$}
	\end{subfigure}
	\begin{subfigure}{.5\textwidth}
		\centering
		\includegraphics[width=\textwidth, height=.4\textheight]{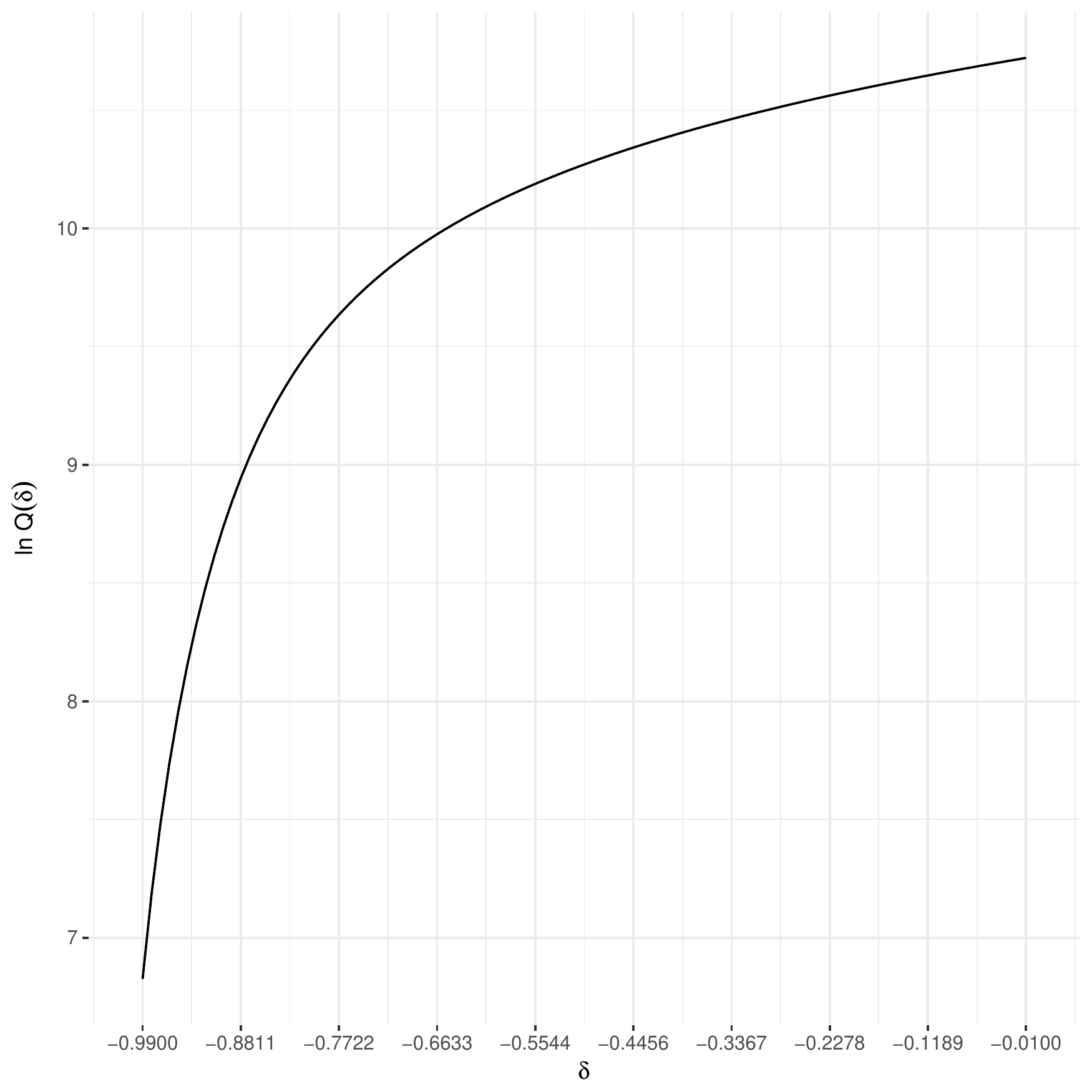}
		\caption{The left half ($-1<\delta<0$) of $\ln Q_{5,0.4}$}
	\end{subfigure}%
	\begin{subfigure}{.5\textwidth}
		\centering
		\includegraphics[width=\textwidth, height=.4\textheight]{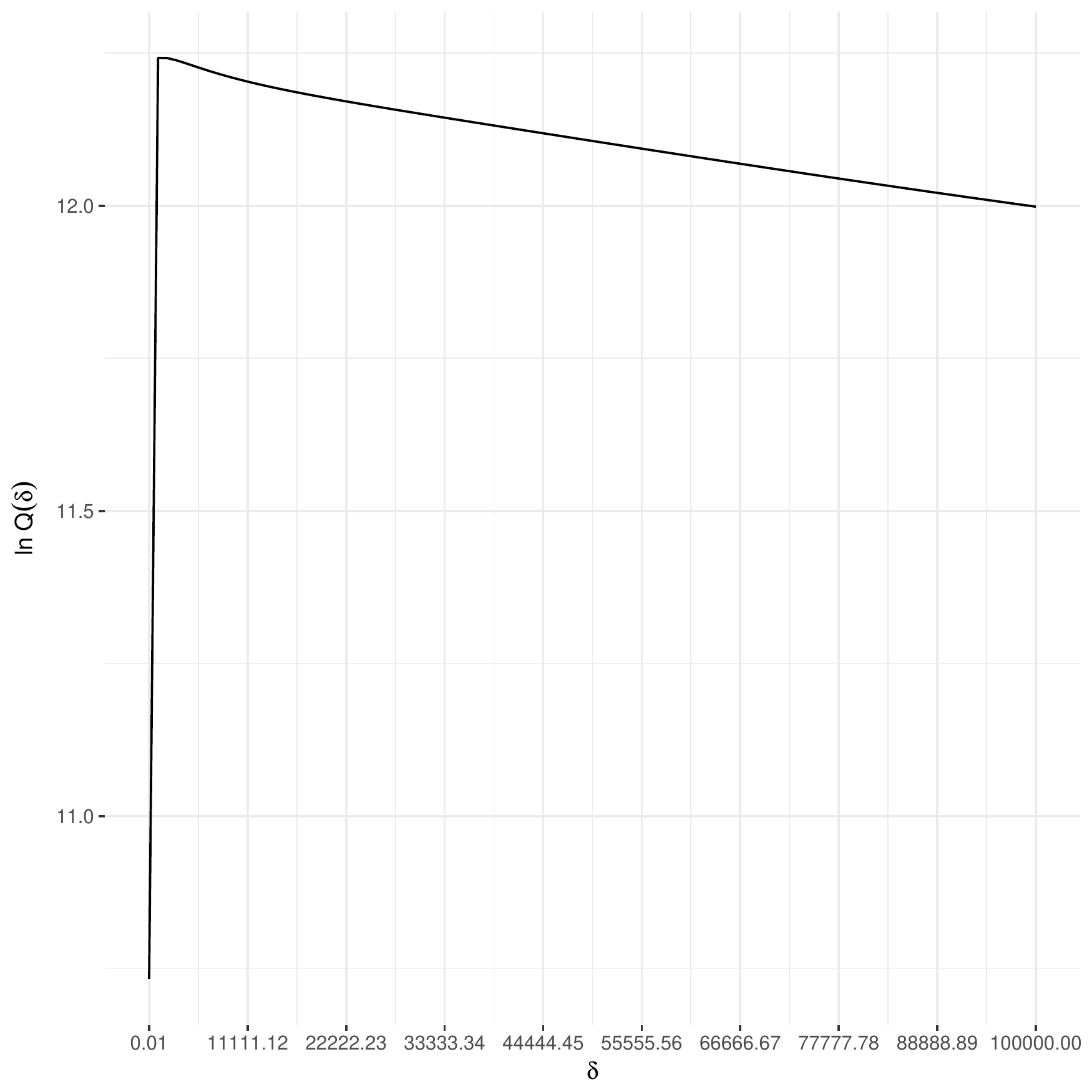}
		\caption{The right half ($\delta>0$) of $\ln Q_{5,0.4}$}
	\end{subfigure}
	\caption{Plots of $\ln Q_{2,0}$ and $\ln Q_{5,0.4}$ for Tecator\texttrademark~data (spectra vs. fat).
		Each pair of curves (i.e., the top two or bottom two) applies to 
		Tecator\texttrademark~data (spectra vs. fat) discussed in \autoref{sec:real.data},
		with two different sets of values for $(p,\alpha)$.
		Neither pair of graphs shows more than one maximizer.
		} 
	\label{fig:curve.Q}
\end{figure}

\autoref{prop:explicit.w.hat} suggests 
merely considering $w$ in the form of \eqref{eq:explicit.w.hat}.
It helps us narrow down the search scope for $\hat{w}_{p,\alpha}$
by reformulating \eqref{eq:wp.fc} into a univariate maximization problem.
The only unknown item in \eqref{eq:explicit.w.hat}, $\hat{\delta}^{(p,\alpha)}$, 
is taken as
\[
	\hat{\delta}^{(p,\alpha)}
	=\argmax_{\delta\in(-1,0)\cup(0,\infty)}Q_{p,\alpha}(\delta)
	=\argmin_{\delta\in(-1,0)\cup(0,\infty)}-\ln Q_{p,\alpha}(\delta),	
\]
where we obtain
\begin{equation}\label{eq:Q}
	\begin{aligned}
		Q_{p,\alpha}(\delta)
		=
		&\left(
			\sum_{j=1}^{\infty}
			\frac{\widehat{\cov}^2\left(
				\widehat{Y}^{(p,\alpha)},
				\int_{\mathcal{T}}\widehat{X}^{(p,\alpha)}\hat{\phi}_j^{(p,\alpha)}
				\right)}
			{\hat{\lambda}_j^{(p,\alpha)}+\hat{\lambda}_1^{(p,\alpha)}/\delta}
		\right)^2 \\
		&\times\left(
			\sum_{j=1}^{\infty}
			\frac{\widehat{\cov}^2\left(
				\widehat{Y}^{(p,\alpha)},
				\int_{\mathcal{T}}\widehat{X}^{(p,\alpha)}\hat{\phi}_j^{(p,\alpha)}
				\right)}
			{\left(\hat{\lambda}_j^{(p,\alpha)}+\hat{\lambda}_1^{(p,\alpha)}/\delta\right)^2}
		\right)^{\frac{\alpha}{1-\alpha}}\\
		&\times\left(
			\sum_{j=1}^{\infty}
			\frac{\widehat{\cov}^2\left(
				\widehat{Y}^{(p,\alpha)},
				\int_{\mathcal{T}}\widehat{X}^{(p,\alpha)}\hat{\phi}_j^{(p,\alpha)}
				\right)
				\widehat{\var}\left(
					\int_{\mathcal{T}}\widehat{X}^{(p,\alpha)}\hat{\phi}_j^{(p,\alpha)}
				\right)}
			{\left(\hat{\lambda}_j^{(p,\alpha)}+\hat{\lambda}_1^{(p,\alpha)}/\delta\right)^2}
		\right)^{\frac{\alpha}{1-\alpha}-1}
	\end{aligned}
\end{equation}
by substituting the right-hand side of \eqref{eq:explicit.w.hat}
for $w$ in \eqref{eq:T.p.star.hat}.
The univariate function $\ln Q_{p,\alpha}$ depends on not only $p$ and $\alpha$ but also observed data,
which makes it inconvenient to theoretically investigate this function's plot.
However,
for the specific datasets to be investigated in \autoref{sec:numerical},
there seems no more than one local maximum within either $(-1,0)$ or $(0,\infty)$;
see \autoref{fig:curve.Q}.
As a result, 
the maximization in each piece
is able to be handled by arbitrary symbolic computation program.

To reduce computational burden and increase the efficiency of \autoref{alg},
we compute
$\widehat{X}_i^{(p,\alpha)}$ and $\hat{\beta}_{p,\alpha}$
in a recursive way,
namely,
for $i=1,\ldots,n$,
\[	
	\widehat{X}_i^{(p,\alpha)} 
	=\widehat{X}_i^{(p-1,\alpha)}-
		\widehat{\var}^{-\frac{1}{2}}\left(
			\int_{\mathcal{T}}X\hat{w}_{p-1,\alpha}
		\right)
		\cdot
		\left(
			\int_{\mathcal{T}}
			(X_i-\bar{X})\hat{w}_{p-1,\alpha}
		\right)
		\cdot
		\widehat{V}_X(\hat{w}_{p-1,\alpha}),
\]
and
\[
	\hat{\beta}_{p,\alpha} 
	=\hat{\beta}_{p-1,\alpha}+
		\widehat{\cov}\left( Y, \int_{\mathcal{T}}X\hat{w}_{p,\alpha} \right)
		\cdot
		\widehat{\var}^{-\frac{1}{2}}\left(
			\int_{\mathcal{T}}X\hat{w}_{p,\alpha}
		\right)
		\cdot
		\hat{w}_{p,\alpha},
\]
starting with 
$\widehat{X}_i^{(1,\alpha)}=X_i-\bar{X}$,
$\widehat{Y}_i^{(1,\alpha)}=Y_i-\bar{Y}$
and $\hat{\beta}_{0,\alpha}=0$.

\subsection{Tuning parameters}\label{subsection:tuning}

The result of functional CR relies on the choice of two parameters:
$\alpha$, the continuum parameter, 
and $p$, the number of basis functions included in the model.
Favoring a much lower expense in computation,
we tune them through the generalized cross-validation (GCV, \citet{CravenWahba1978}).
For each possible pair $(p,\alpha)$,
a GCV-type criterion employed here is 
\[
	\text{GCV}(p,\alpha)=
	\frac{\sum_{i=1}^n\left(Y_i-\hat{\eta}_{p,\alpha}(X_i)\right)^2}{(n-p-1)^2}.
\]
The minimizer of $\text{GCV}(p,\alpha)$ is chosen as the optimal combination;
see \autoref{alg} for details.

\begin{algorithm}
	\caption{Functional CR tuned by GCV}
	\label{alg}
	\begin{algorithmic}[]
		\State $pmax\leftarrow$ number of basis functions selected by functional PCR.
		\For {$\alpha$ in a finite set and $p$ from 1 to $pmax$} 
			\For {$i$ from 1 to $n$} 
				\If {$p=1$}
					\State $\widehat{X}_i^{(p,\alpha)}\leftarrow X_i-\bar{X}$.
					\State $\widehat{Y}_i^{(p,\alpha)}\leftarrow Y_i-\bar{Y}$.
					\State $\hat{\beta}_{p-1,\alpha}\leftarrow 0$.
				\Else
					\State $\widehat{X}_i^{(p,\alpha)}\leftarrow 
						\widehat{X}_i^{(p-1,\alpha)}-
						c_2\cdot c_3\cdot
						\widehat{V}_X(\hat{w}_{p-1,\alpha})$.
					\State $\widehat{Y}_i^{(p,\alpha)}\leftarrow 
						\widehat{Y}_i^{(p-1,\alpha)}-\hat{\eta}_{p-1,\alpha}(X_i)$.
				\EndIf
			\EndFor
			\State 
				$\hat{\lambda}_j^{(p,\alpha)},\hat{\phi}_j^{(p,\alpha)}\leftarrow$ 
					the $j$-th eigenvalue and eigenfunction of
				 	$\widehat{V}_{\widehat{X}^{(p,\alpha)}}$.
			\State 
				$a_j\leftarrow
					\widehat{\cov}\left(
						\widehat{Y}^{(p,\alpha)},
						\int_{\mathcal{T}}\widehat{X}^{(p,\alpha)}\hat{\phi}_j^{(p,\alpha)}
					\right)$.
			\State
				$b_j\leftarrow
					\widehat{\var}\left(
					\int_{\mathcal{T}}\widehat{X}^{(p,\alpha)}\hat{\phi}_j^{(p,\alpha)}
					\right)$.
			\State $Q_{p,\alpha}(\delta)\leftarrow
				\left( \sum_{j=1}^{\infty}\frac{a_j^2}{\hat{\lambda}_j^{(p,\alpha)}+\hat{\lambda}_1^{(p,\alpha)}/\delta} \right)^2
				\left( 
					\sum_{j=1}^{\infty}
					\frac{a_j^2}{ \left(\hat{\lambda}_j^{(p,\alpha)}+\hat{\lambda}_1^{(p,\alpha)}/\delta \right)^2} 
				\right)^{\frac{\alpha}{1-\alpha}}
				\left( 
					\sum_{j=1}^{\infty}
					\frac{a_j^2b_j}{\left(\hat{\lambda}_j^{(p,\alpha)}+\hat{\lambda}_1^{(p,\alpha)}/\delta\right)^2}
				\right)^{\frac{\alpha}{1-\alpha}-1}$.
			\State $\hat{\delta}^{(p,\alpha)}\leftarrow
				\argmin_{\delta\in(-1,0)\cup(0,\infty)}-\ln Q_{p,\alpha}(\delta)$.
			\State $\hat{w}_{p,\alpha}\leftarrow
				\left( 
					\sum_{j=1}^{\infty}
					\frac{a_j^2}{\left( \hat{\lambda}_j^{(p,\alpha)}+\hat{\lambda}_1^{(p,\alpha)}/\hat{\delta}^{(p,\alpha)} \right)^2} 
				\right)^{-\frac{1}{2}}
					\sum_{j=1}^{\infty}
					\frac{a_j}{\hat{\lambda}_j^{(p,\alpha)}+\hat{\lambda}_1^{(p,\alpha)}/\hat{\delta}^{(p,\alpha)}}
					\hat{\phi}_j^{(p,\alpha)}$.
			\State 
				$c_1\leftarrow
					\widehat{\cov}\left( 
					Y, \int_{\mathcal{T}}X\hat{w}_{p,\alpha}
					\right)$.
			\State
				$c_2\leftarrow
					\widehat{\var}^{-\frac{1}{2}}\left(
					\int_{\mathcal{T}}X\hat{w}_{p,\alpha}
					\right)$.
			\State
				$c_3\leftarrow
					\int_{\mathcal{T}}
					\widehat{X}_i^{(1,\alpha)}\hat{w}_{p,\alpha}$.
			\State 
				$\hat{\beta}_{p,\alpha}\leftarrow
					\hat{\beta}_{p-1,\alpha} 
					+c_1\cdot c_2\cdot \hat{w}_{p,\alpha}$.
			\For {$i$ from 1 to $n$}
				\State $\hat{\eta}_{p,\alpha}(X_i)\leftarrow
					\bar{Y}+\int_{\mathcal{T}}
					\widehat{X}_i^{(1,\alpha)}
					\hat{\beta}_{p,\alpha}$.
			\EndFor
			\State $\text{GCV}(p,\alpha)\leftarrow
				\sum_{i=1}^n\left(Y_i-\hat{\eta}_{p,\alpha}(X_i)\right)^2/(n-p-1)^2$.
		\EndFor
		\State $(p,\alpha)_{\text{optimal}}\leftarrow
						\underset{(p,\alpha)}{\argmin}\text{GCV}(p,\alpha)$.
	\end{algorithmic}
\end{algorithm}

\section{Numerical illustration}\label{sec:numerical}

To illustrate the performance of functional CR,
the result given by our method is compared with those from
supervised FPCA (\citet{NieWangLiuCao2018}),
pFPLS (\citet{AguileraAguilera-MorilloPreda2016}),
FPLS$_R$-, FPCR$_R$-REML 
(both recommended by \citet{ReissOgden2007} after a series of comparisons)
and smoothed functional PCA 
(\citet[Section 9.3]{RamsaySilverman2005}).
Among these the first four are supervised,
while the other two are categorized as unsupervised.

With the aid of R (\citet{R}), RStudio\texttrademark\ (\citet{RStudio}) and R-package \texttt{fda} (\citet{R-fda}),
we code all the methods mentioned in the preceding paragraph
except for FPCR$_R$-REML  
(implemented by R-function \texttt{fpcr} 
coded by its proposers and included in R-package \texttt{refund} 
jointly created by \citet{R-refund}).

\subsection{Simulation study}\label{sec:simulation}

The dataset \texttt{CanadianWeather} in \citet{R-fda}
contains the (base 10 logarithm of) precipitation at 35 different locations in Canada averaged over 1960 to 1994.
We extract the mean function $\mu_X$ and
the top $j$-th eigenvalue $\lambda_j$ and eigenfunction $w_{j,\text{FPC}}$ of the covariance operator, $j=1,2,3$,
from this dataset.
Each sample in this simulation consists of 35 artificial curves on $\mathcal{T}=[1,365]$ of the following form:
\[
	X_i=\mu_X + \sum_{j=1}^3 \xi_{ij}w_{j,\text{FPC}},
	\quad i=1,\ldots,35,
\]
where $\xi_{ij}$ follows $N(0,\lambda_j)$ independently
and 35 is the number of curves included in \texttt{CanadianWeather}.
Further,
$Y_i$ are generated as
\[
	Y_i =  \int_{\mathcal{T}}\beta X_i + \varepsilon_i,
	\quad i=1,\ldots,35,
\]
with $\varepsilon_i\overset{\text{iid}}{\sim} N(0,\sigma)$.
The quantity $\int_{\mathcal{T}}\mu_X\beta/\sigma$ is informally referred to as the signal-to-noise-ratio (SNR).

We consider two sorts of coefficient function coupled with three levels of SNR (2,10 and 20):
\begin{enumerate}[label=(\roman*)] 
	\item $\beta=w_{1,\text{FPC}}$; \label{case:1st.eigen}
	\item $\beta=w_{3,\text{FPC}}$. \label{case:3rd.eigen}
\end{enumerate}
No matter how supervised they are, 
all the methods are applicable to Scenario \ref{case:1st.eigen},
while Scenario \ref{case:3rd.eigen} is to imitate the situation
where the coefficient function is orthogonal to the top few eigenfunctions of $V_X$,
i.e., the initial target our proposal is shooting at.
For each combination of $\beta$ and SNR,
we generate 200 samples and apply the six techniques to each dataset to estimate $\beta$.
The estimation quality is directly measured by the square root of mean squared errors (RMSE) (on each $t\in\mathcal{T}$)
of estimated coefficient functions.

The candidate pool of $(p, \alpha)$ for functional CR is a $2\times 11$ grid,
$\{1,2\}\times \{0,.1,.2,.3,.4,.5,.6,.7,.8,.9,.999\}$,
where the scope of $p$,
$\{1,2\}$,
remains for all five other methods.
This setting is reasonable because the first two functional principal components capture around 97\% of the total variation in predictor.
In the implementation of smoothed functional PCA, supervised FPCA and pFPLS,
smoothing penalty parameters are chosen from $\{0,1,10,10^2,10^3,10^4,10^5\}$.
Moreover,
as suggested by \citet{NieWangLiuCao2018},
candidate values of the ``weight'' parameter needed by supervised FPCA are taken from
$\{0,.1,.2,.3,.4,.5,.6,.7,.8,.9,1\}$.

\begin{figure}
	\begin{subfigure}{\textwidth}
		\centering
		\includegraphics[width=.8\textwidth, height=.28\textheight]{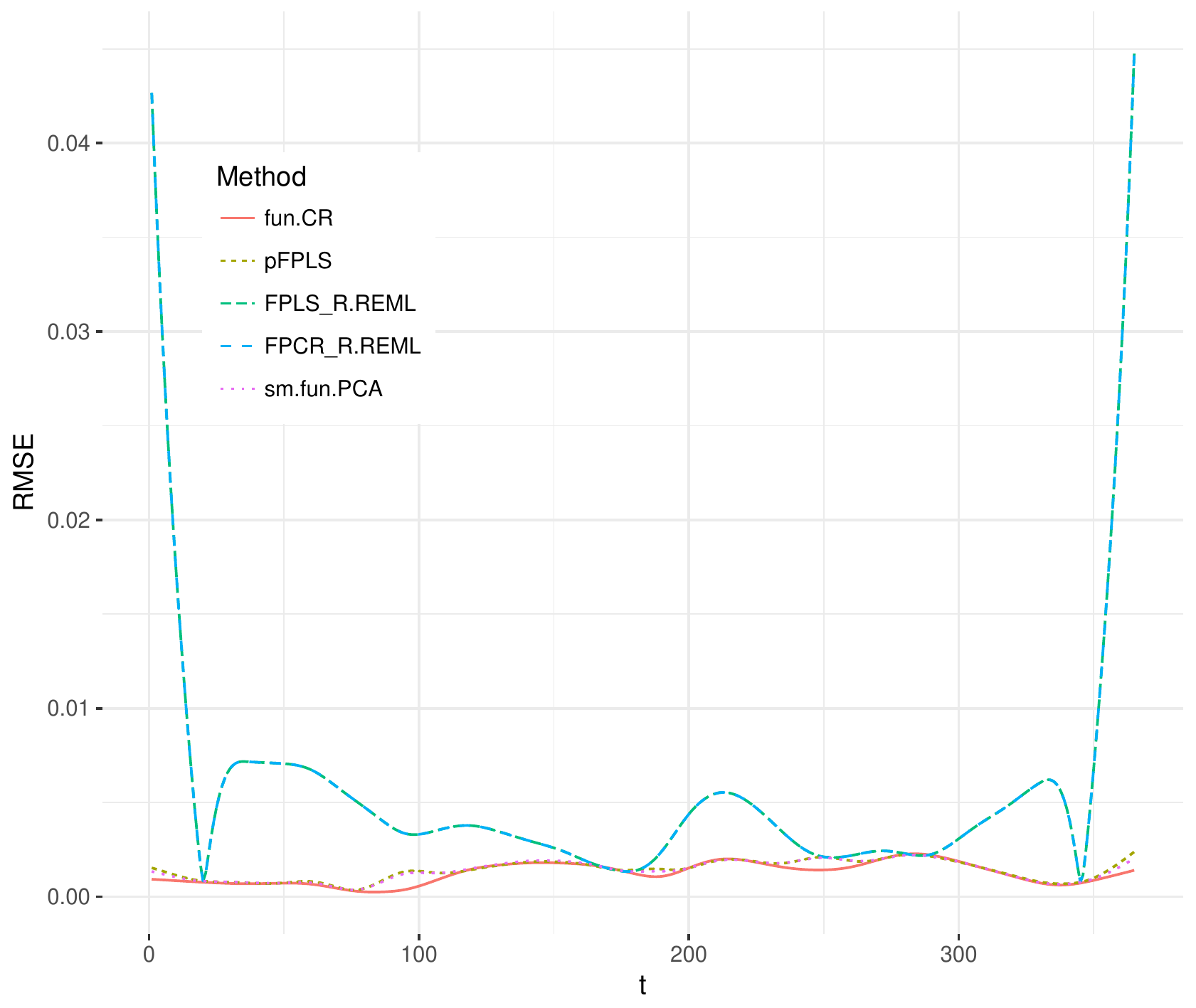}
		\caption{SNR=20}
	\end{subfigure}
	
	\begin{subfigure}{\textwidth}
		\centering
		\includegraphics[width=.8\textwidth, height=.28\textheight]{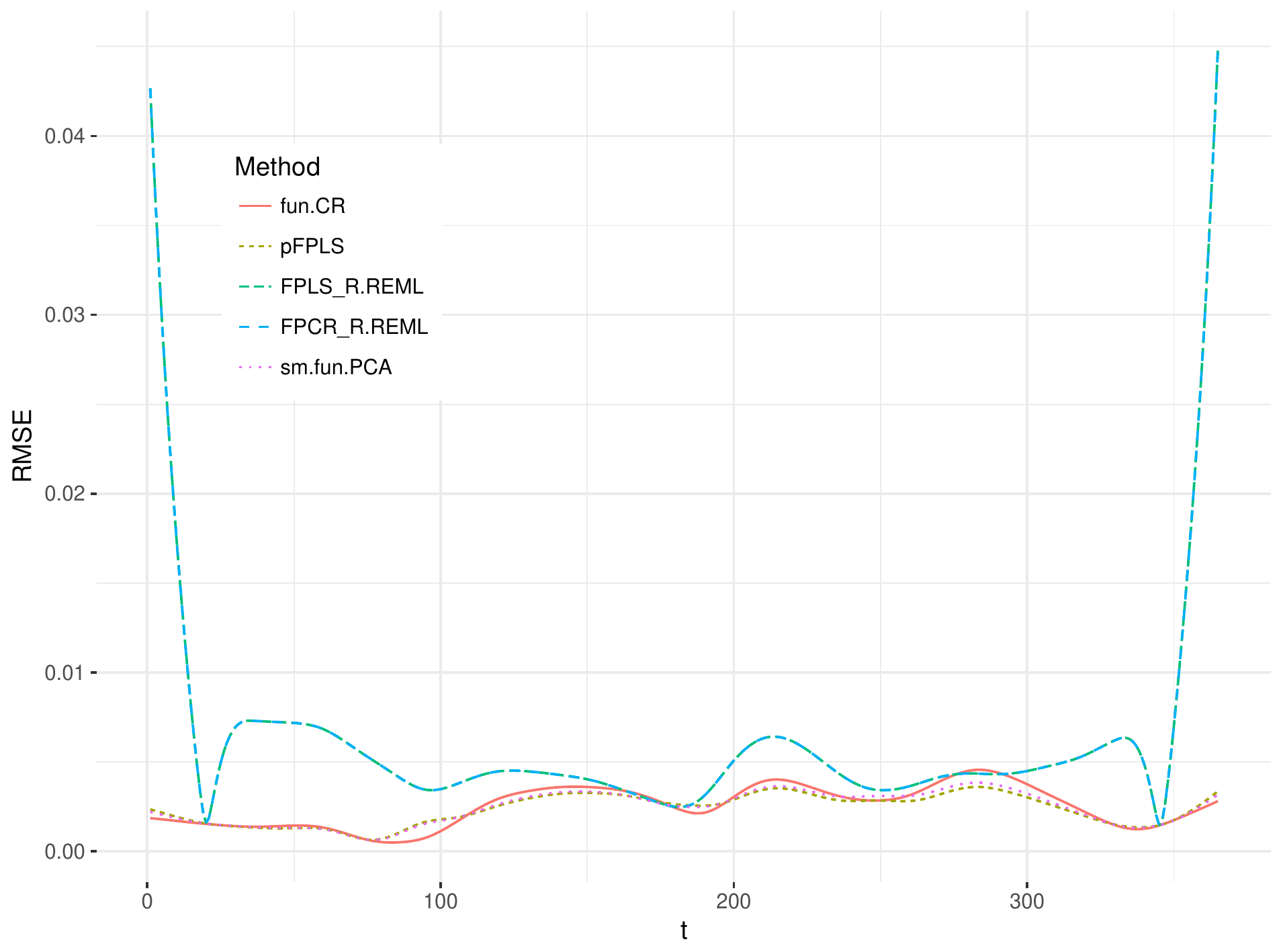}
		\caption{SNR=10}
	\end{subfigure}%
	
	\begin{subfigure}{\textwidth}
		\centering
		\includegraphics[width=.8\textwidth, height=.28\textheight]{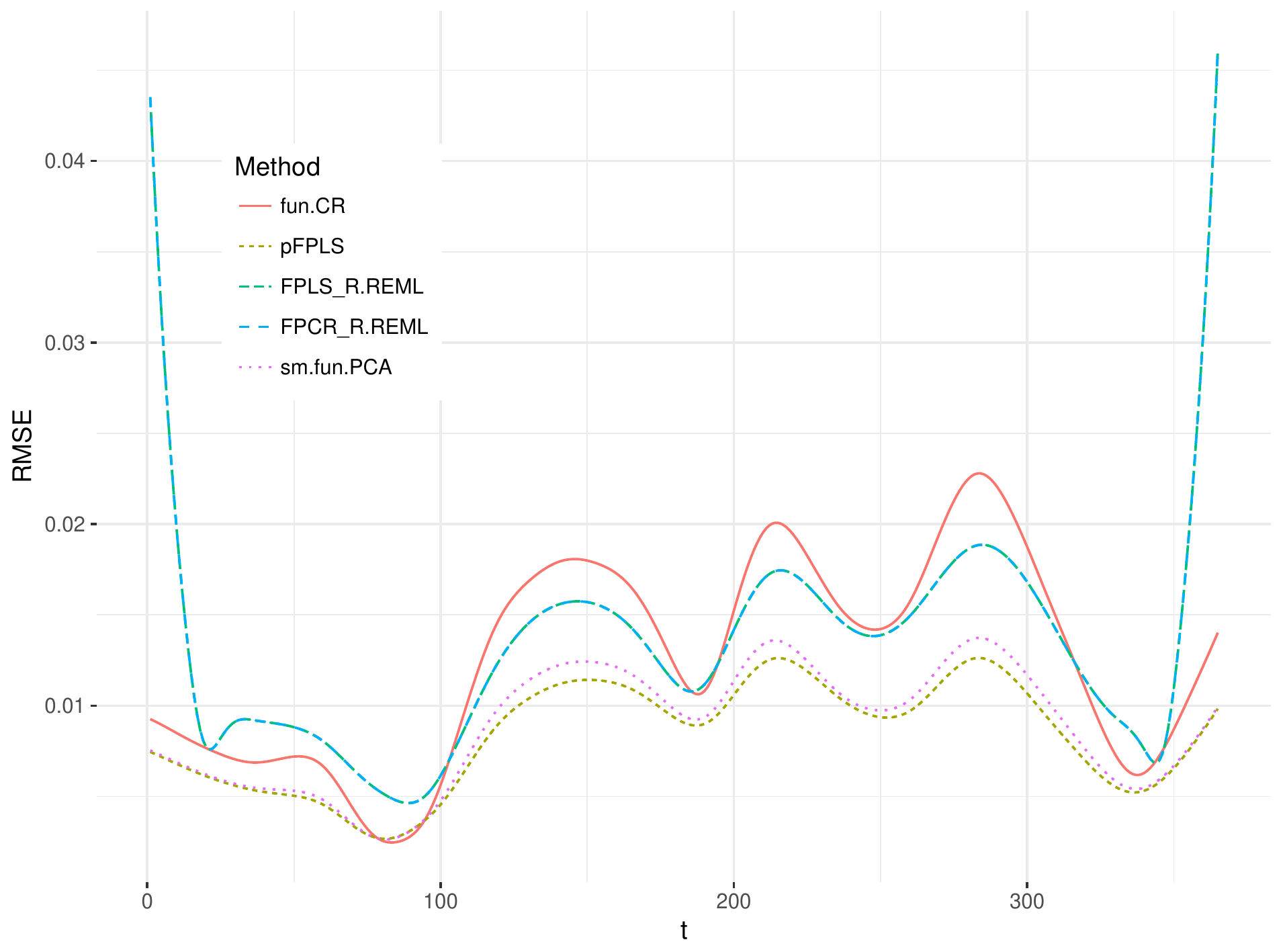}
		\caption{SNR=2}
	\end{subfigure}%
	
	\caption{
		Plots of RMSE of estimated coefficient function in \textbf{Scenario \ref{case:1st.eigen} of the simulation} ($\beta=w_{1,\text{FPC}}$) 
		with different SNRs.
		In the legend of each subfigure,
		the five linetypes (or colors),
		from top to bottom,
		correspond to functional CR, 
		pFPLS, FPLS$_R$-REML, FPCR$_R$-REML and smoothed functional PCA, respectively.
		Curves for FPLS$_R$-REML and FPCR$_R$-REML almost overlap each other.
		Supervised FPCA does not perform well in estimation for this case 
		and hence its RMSE curve is not shown.
	} 
	\label{fig:rmse.simu1}
\end{figure}

When $\beta=w_{1,\text{FPC}}$ and SNR is more than moderate ($\text{SNR}=10,20$), 
functional CR (solid red) performs as well as pFPLS and smoothed functional PCA
and slightly better than FPLS$_R$-REML and FPCR$_R$-REML
whose RMSEs become dramatically high at the two extremes of the domain.
For any method,
RMSEs are enlarged with a decrease of SNR, 
while functional CR (solid red) seems more sensitive to the change of SNR;
see \autoref{fig:rmse.simu1}.

\begin{figure}
	\begin{subfigure}{\textwidth}
		\centering
		\includegraphics[width=.8\textwidth, height=.28\textheight]{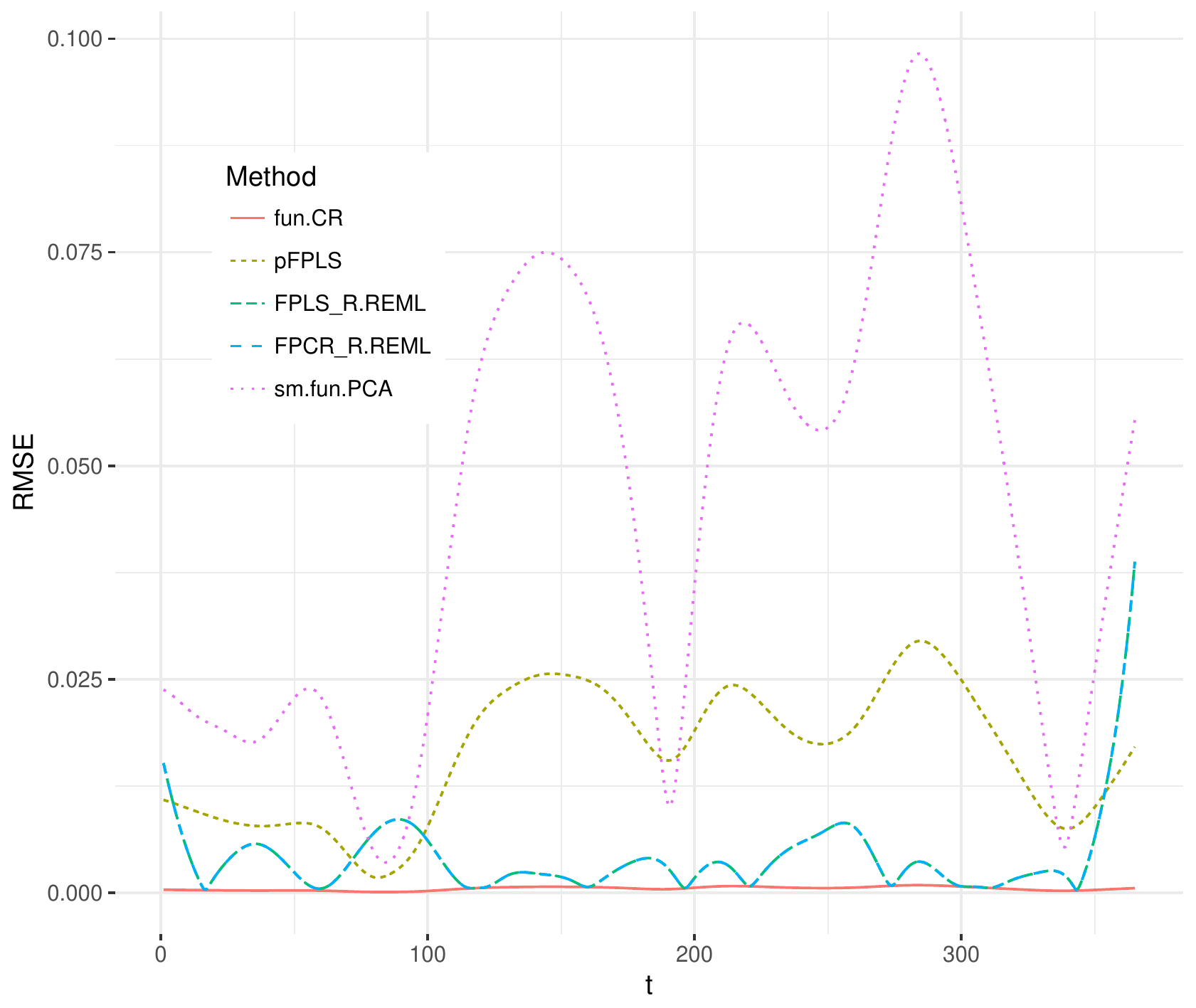}
		\caption{SNR=20}
	\end{subfigure}
	
	\begin{subfigure}{\textwidth}
		\centering
		\includegraphics[width=.8\textwidth, height=.28\textheight]{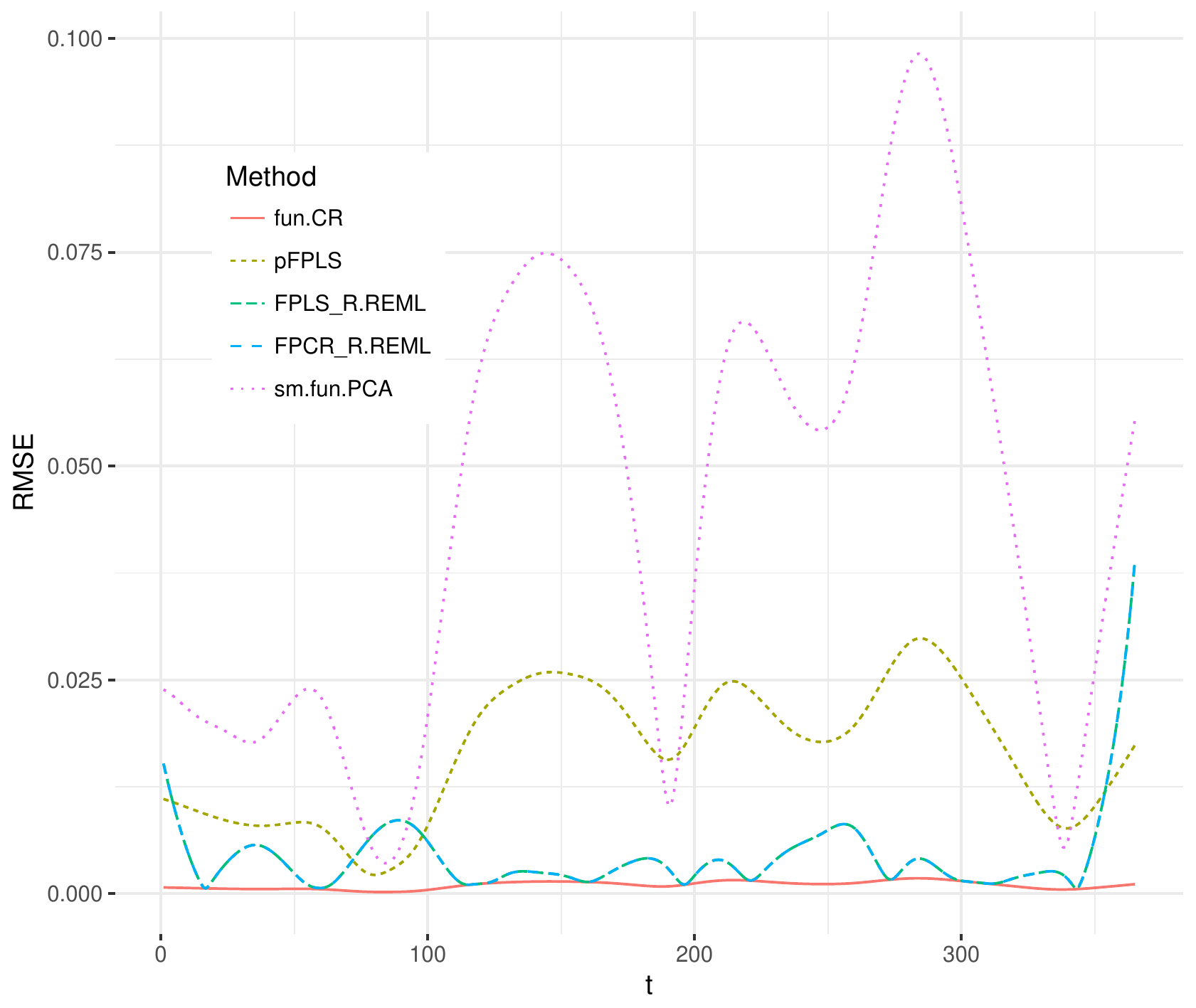}
		\caption{SNR=10}
	\end{subfigure}%
	
	\begin{subfigure}{\textwidth}
		\centering
		\includegraphics[width=.8\textwidth, height=.28\textheight]{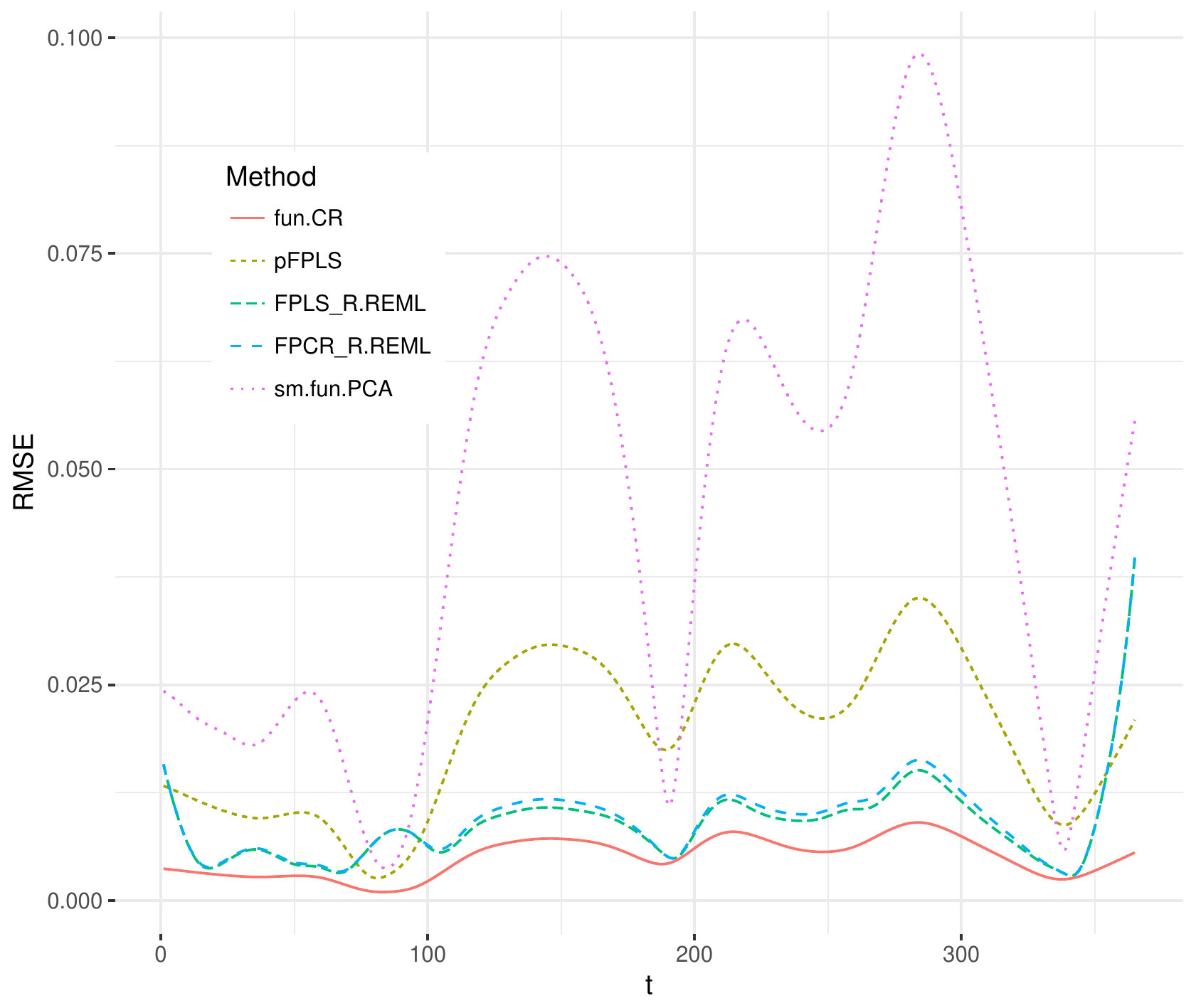}
		\caption{SNR=2}
	\end{subfigure}%

	\caption{
		Plots of RMSE of estimated coefficient functions 
		in \textbf{Scenario \ref{case:3rd.eigen} of the simulation} ($\beta=w_{3,\text{FPC}}$) 
		with different SNRs.
		In the legend of each subfigure,
		the five linetypes (or colors),
		from top to bottom,
		correspond to functional CR, 
		pFPLS, FPLS$_R$-REML, FPCR$_R$-REML and smoothed functional PCA, respectively.
		Curves for FPLS$_R$-REML and FPCR$_R$-REML almost overlap each other.
		Supervised FPCA does not perform well in estimation for this case 
		and hence its RMSE curve is not shown.	
	}
	\label{fig:rmse.simu2}
\end{figure}

Unsurprisingly, 
as shown in \autoref{fig:rmse.simu2},
Scenario \ref{case:3rd.eigen} ($\beta=w_{3,\text{FPC}}$) does not favor the smoothed functional PCA
which involves only two basis functions nearly orthogonal to $\beta$.
Oppositely,
functional CR (solid red) outperforms competitors regardless of $\text{SNR}$
and returns the lowest RMSE at almost every $t\in\mathcal{T}$,
in spite of the setting violating the assumption of \autoref{prop:explicit.w}.
Note that curves for supervised FPCA are never included in Figures \ref{fig:rmse.simu1} and \ref{fig:rmse.simu2}
as RMSEs from supervised FPCA are much larger than those from other approaches; 
either the estimators from \citet{NieWangLiuCao2018} are not consistent or
it may need a larger sample size to reach a more satisfying accuracy. 

\subsection{Application to real data}\label{sec:real.data}

For each of following two datasets,
we randomly take roughly 10\% of all the samples of each dataset 
for testing and the remaining for training.
Repeat the random split for 200 times.
To alleviate impacts from different testing sets
and facilitate the comparison in prediction,
define the relative mean squared prediction error (ReMSPE):
\[
	\text{ReMSPE}=
	\frac{\sum_{i\in I_{\text{test}}}\left(Y_i-\widehat{Y}_i\right)^2}
		{\sum_{i\in I_{\text{test}}}\left(Y_i-\sum_{i\in I_{\text{train}}}Y_i/\#I_{\text{train}}\right)^2},
\]
where $I_{\text{train}}$ and $I_{\text{test}}$ are respective index sets for training and testing data,
$\# I_{\text{train}}$ is the cardinality of $I_{\text{train}}$,
and $\widehat{Y}_i$ is the prediction corresponding to $Y_i$.
For each approach,
generate a boxplot of the 200 ReMSPEs.
As for the candidate pool for tuning parameters, 
we keep all the settings in \autoref{sec:simulation} except the one for $p$;
we raise its upper bound from 2 to 5 
to accommodate the new datasets.

\subsubsection{Medfly data}

Investigated in substantial literature
(see, e.g., \citet{MullerWangCapraLiedoCarey1997} and \citet{SangWangCao2017}), 
the Mediterranean fruit fly, or medfly for short, 
has become indeed a popular object of study,
partly owing to its short lifespan.
Posted at \url{http://faculty.bscb.cornell.edu/~hooker/FDA2008/medfly.Rdata},
the medfly data here records lifespans of 50 female flies 
as well as numbers of eggs laid by each of them in each of the 26 days.
People would like to uncover how lifespan is influenced by fecundity as time goes on
by bridging the curves of egg count to lifespans.

\begin{figure}
	\centering
	\begin{subfigure}{.5\textwidth}
		\centering
		\includegraphics[width=\textwidth, height=.4\textheight]{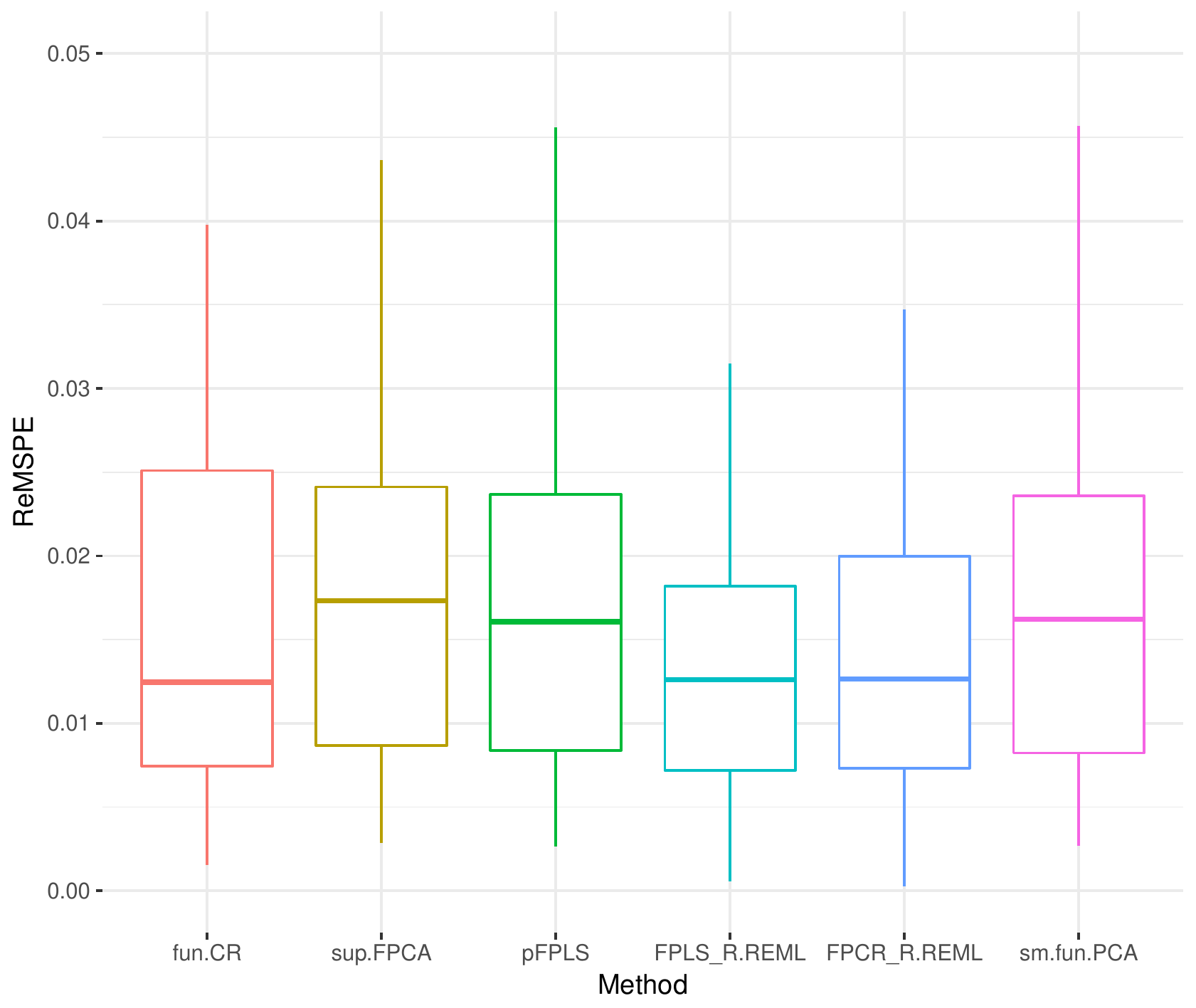}
		\caption{First 50 repeats}
	\end{subfigure}%
	\begin{subfigure}{.5\textwidth}
		\centering
		\includegraphics[width=\textwidth, height=.4\textheight]{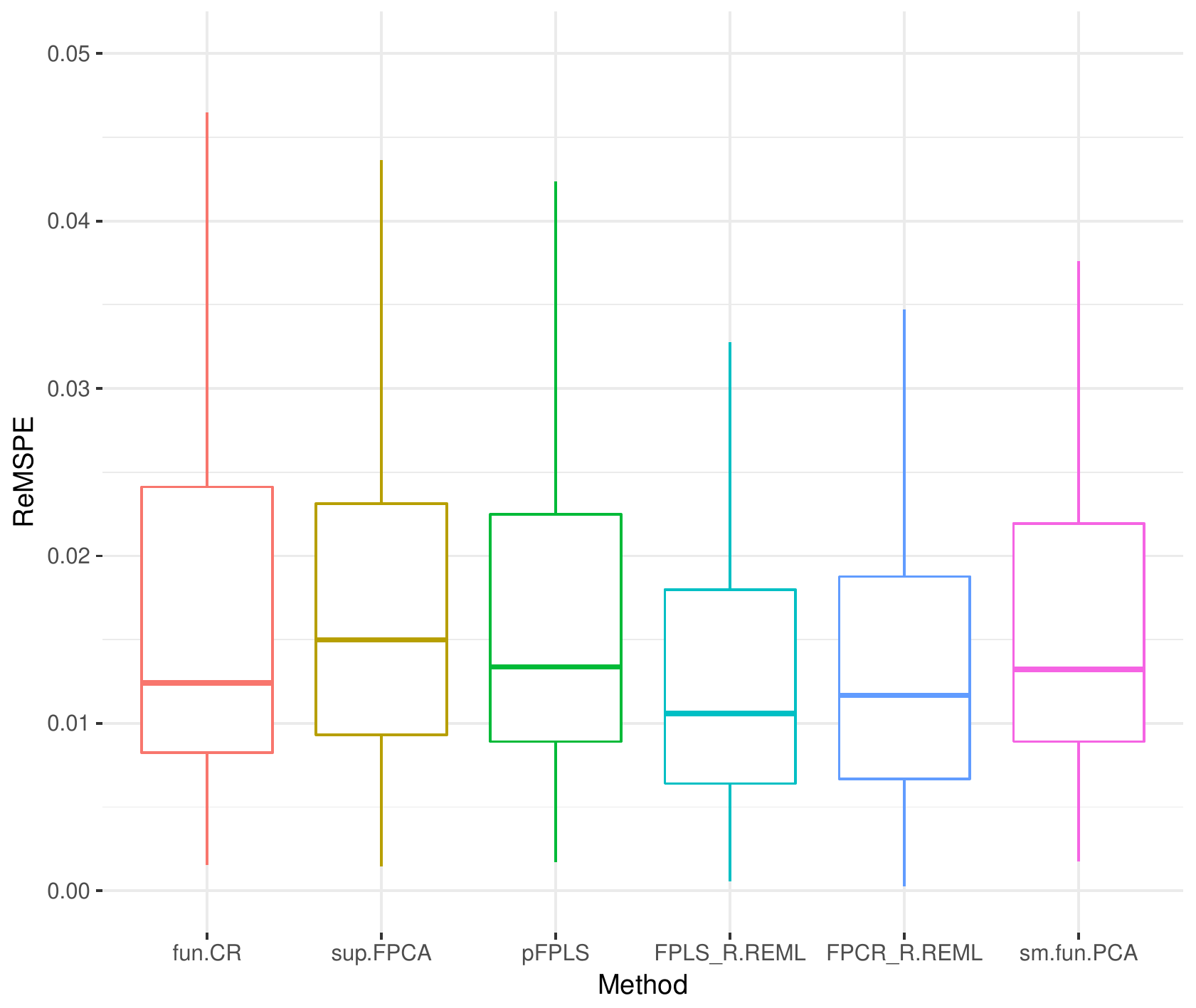}
		\caption{First 200 repeats}
	\end{subfigure}
	\caption{
		Boxplots of ReMSPEs of six methods for \textbf{medfly data}.
		In each subfigure,
		the six boxplots,
		from left to right,
		correspond to functional CR, 
		supervised FPCA, pFPLS, FPLS$_R$-REML, FPCR$_R$-REML and smoothed functional PCA, respectively.
	} 
	\label{fig:boxplot.medfly}
\end{figure}

Taking the egg count and lifespan as predictor and response respectively,
all the six methods,
no matter whether supervised or unsupervised, 
perform fairly close to each,
though FPCR$_R$- and FPLS$_R$-REML appear way better than the other four;
see \autoref{fig:boxplot.medfly}.


\subsubsection{Tecator\texttrademark~data}

A Tecator\texttrademark~Infratec Food and Feed Analyzer
recorded near infrared absorbance spectra 
(ranging from 850 to 1050 nm and divided into 100 channels) 
of 240 finely chopped pure meat samples with different fat, moisture and protein contents.
The dataset is now publicly accessible at \url{http://lib.stat.cmu.edu/datasets/tecator},
containing the absorbance spectra (i.e., the logarithm to base 10 of transmittance at each wavelength)
and the three contents measured in percent by analytic chemistry.

\begin{figure}
	\centering
	\begin{subfigure}{.5\textwidth}
		\centering
		\includegraphics[width=\textwidth, height=.4\textheight]{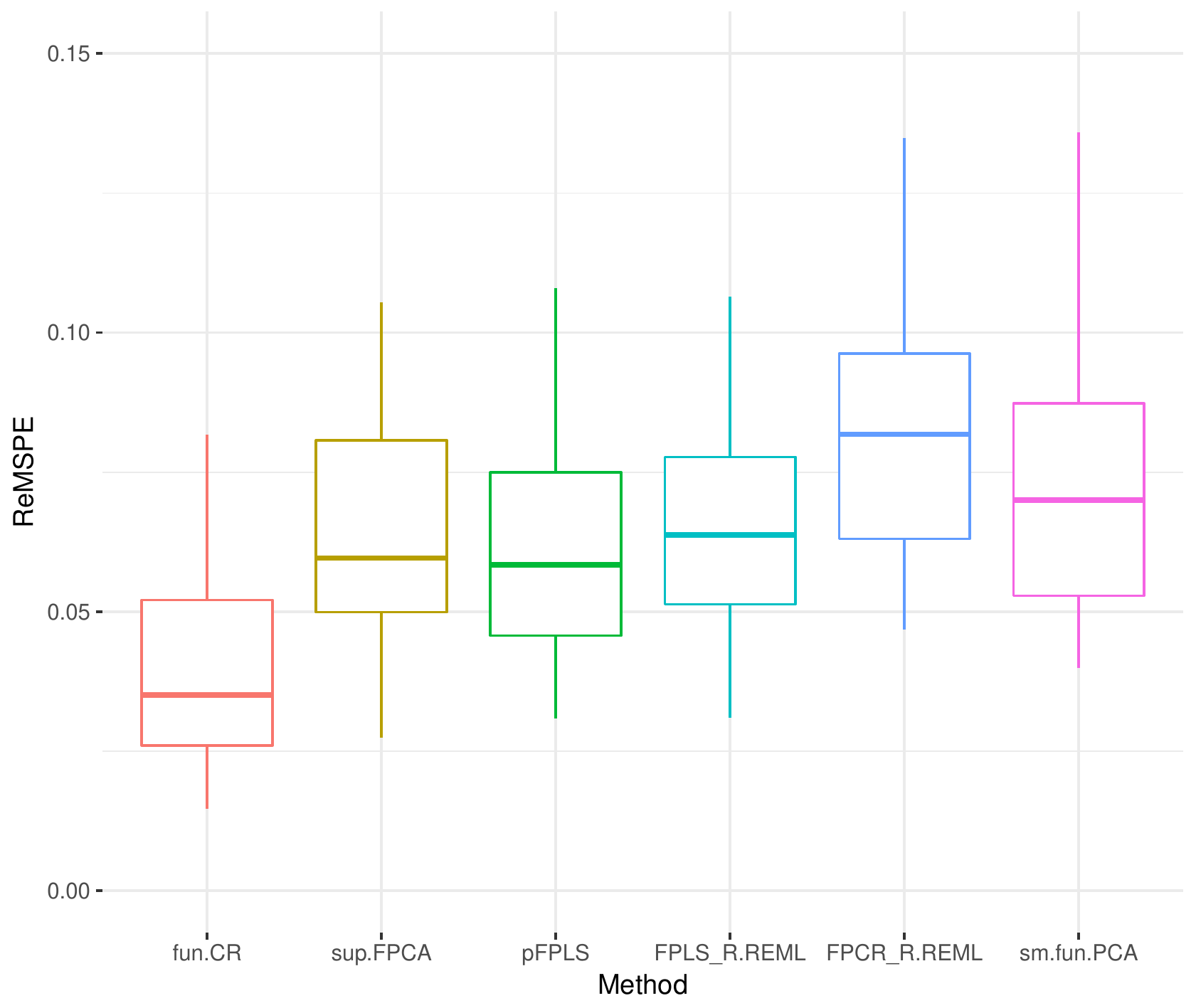}
		\caption{First 50 repeats}
	\end{subfigure}%
	\begin{subfigure}{.5\textwidth}
		\centering
		\includegraphics[width=\textwidth, height=.4\textheight]{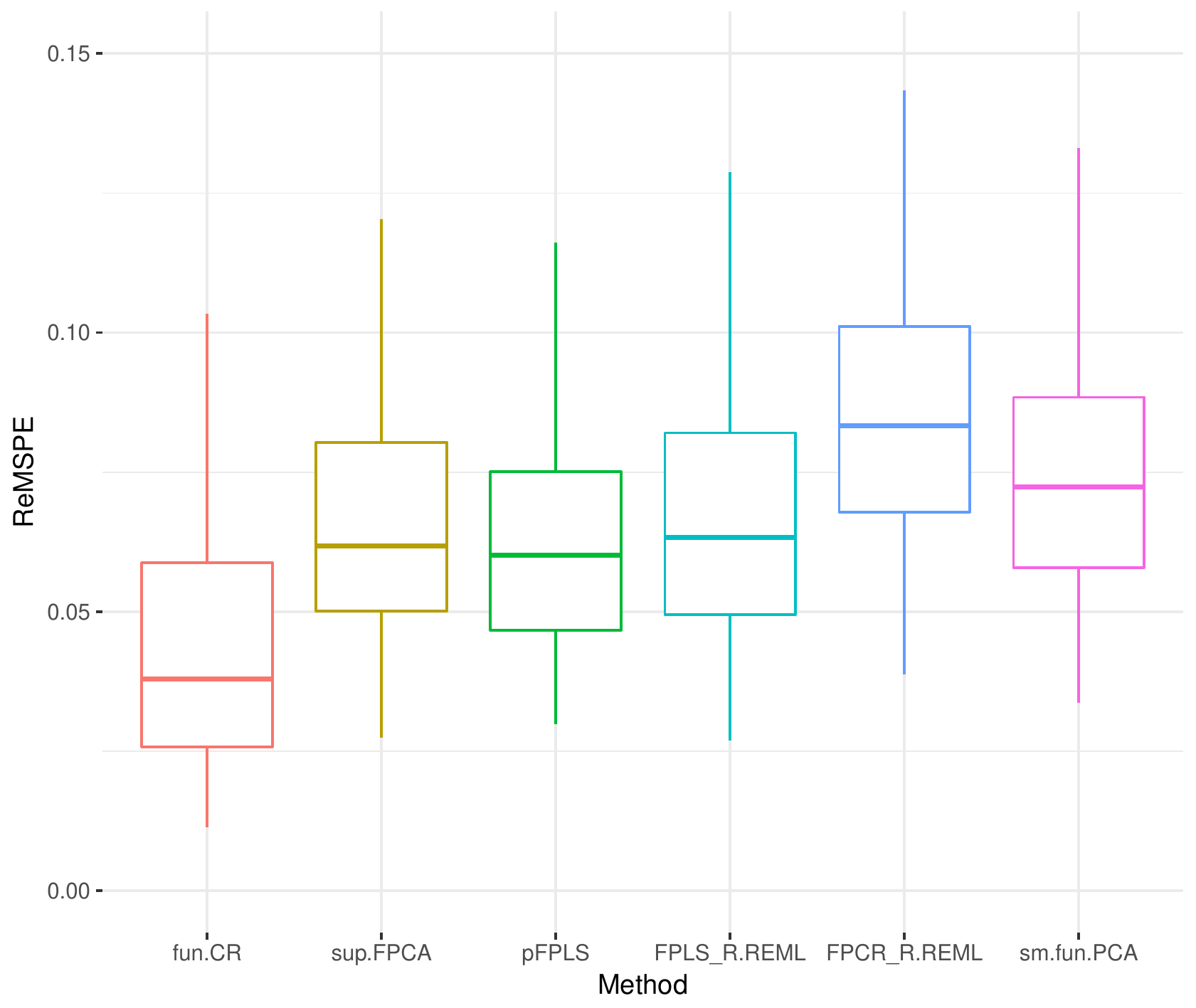}
		\caption{First 200 repeats}
	\end{subfigure}
	\caption{Boxplots of ReMSPEs of six methods for \textbf{Tecator\texttrademark~data (spectra vs. fat)}.
		In each subfigure,
		the six boxplots,
		from left to right,
		correspond to functional CR, supervised FPCA, pFPLS, FPLS$_R$-REML, FPCR$_R$-REML and smoothed functional PCA, respectively.
	} 
	\label{fig:boxplot.fat}
\end{figure}

\begin{figure}
	\centering
	\begin{subfigure}{.5\textwidth}
		\centering
		\includegraphics[width=\textwidth, height=.4\textheight]{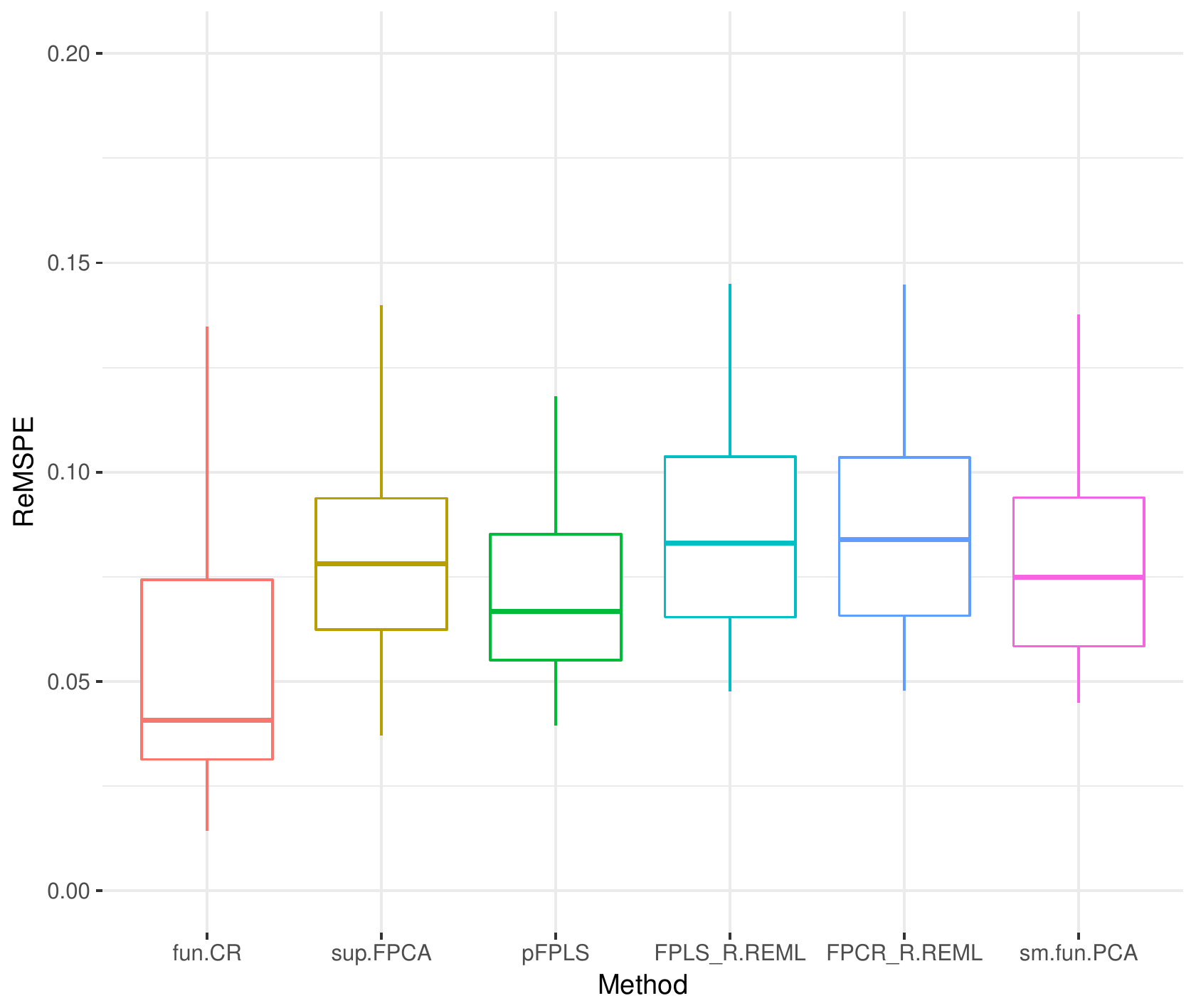}
		\caption{First 50 repeats}
	\end{subfigure}%
	\begin{subfigure}{.5\textwidth}
		\centering
		\includegraphics[width=\textwidth, height=.4\textheight]{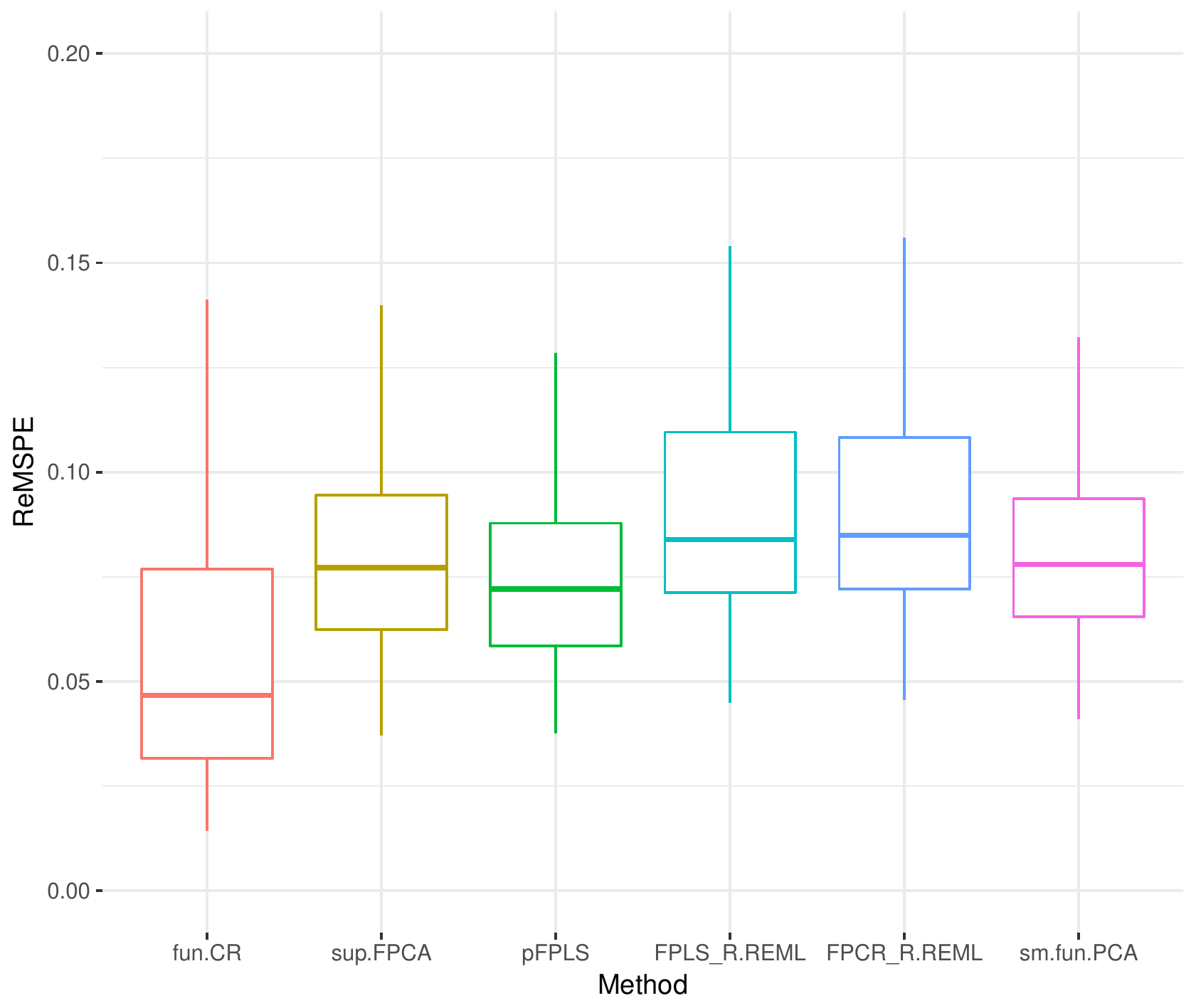}
		\caption{First 200 repeats}
	\end{subfigure}
	\caption{Boxplots of ReMSPEs of six methods for \textbf{Tecator\texttrademark~data (spectra vs. moisture)}.
		In each subfigure,
		the six boxplots,
		from left to right,
		correspond to functional CR, supervised FPCA, pFPLS, FPLS$_R$-REML, FPCR$_R$-REML and smoothed functional PCA, respectively.
	} 
	\label{fig:boxplot.moisture}
\end{figure}

\begin{figure}
	\centering
	\begin{subfigure}{.5\textwidth}
		\centering
		\includegraphics[width=\textwidth, height=.4\textheight]{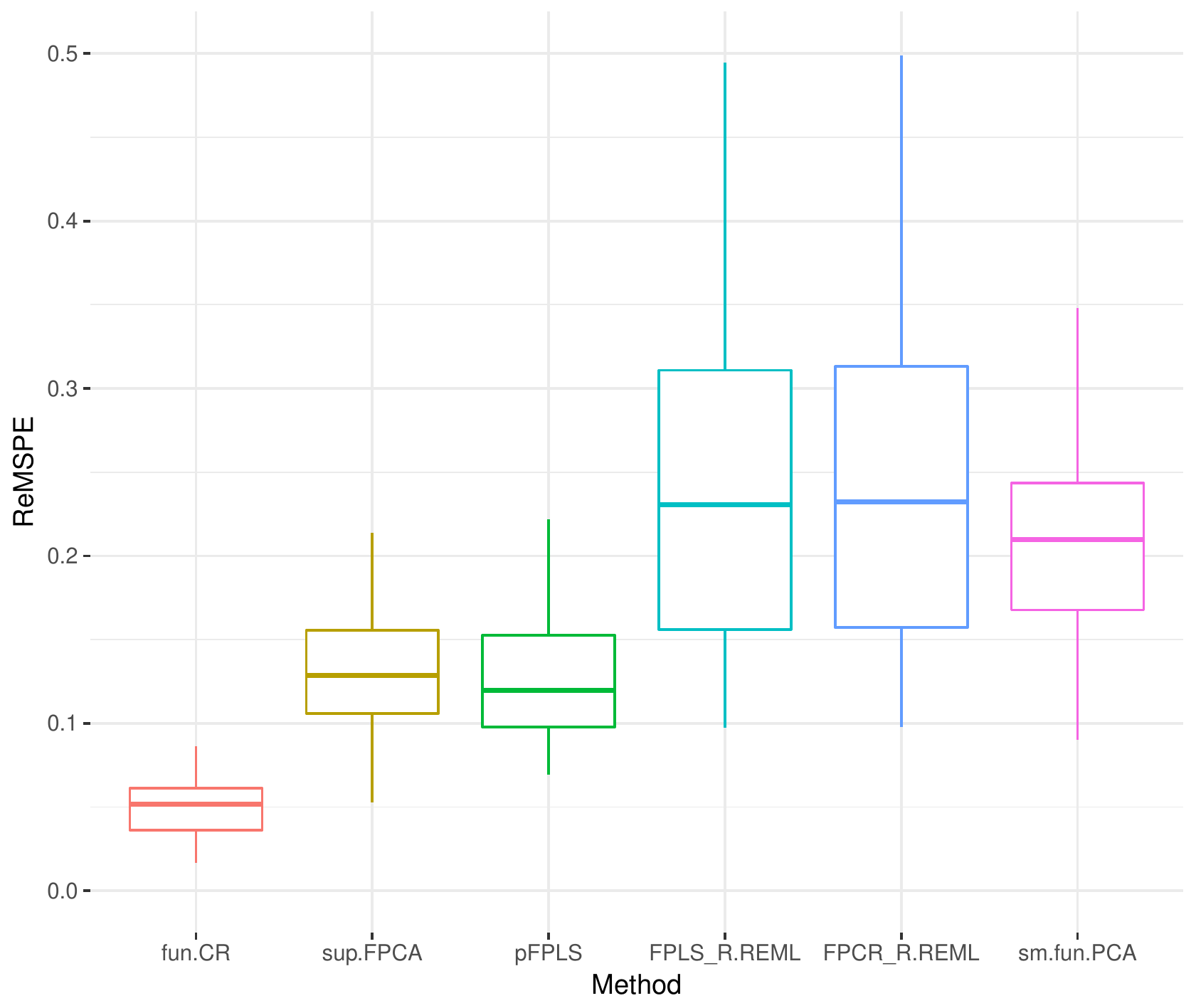}
		\caption{First 50 repeats}
	\end{subfigure}%
	\begin{subfigure}{.5\textwidth}
		\centering
		\includegraphics[width=\textwidth, height=.4\textheight]{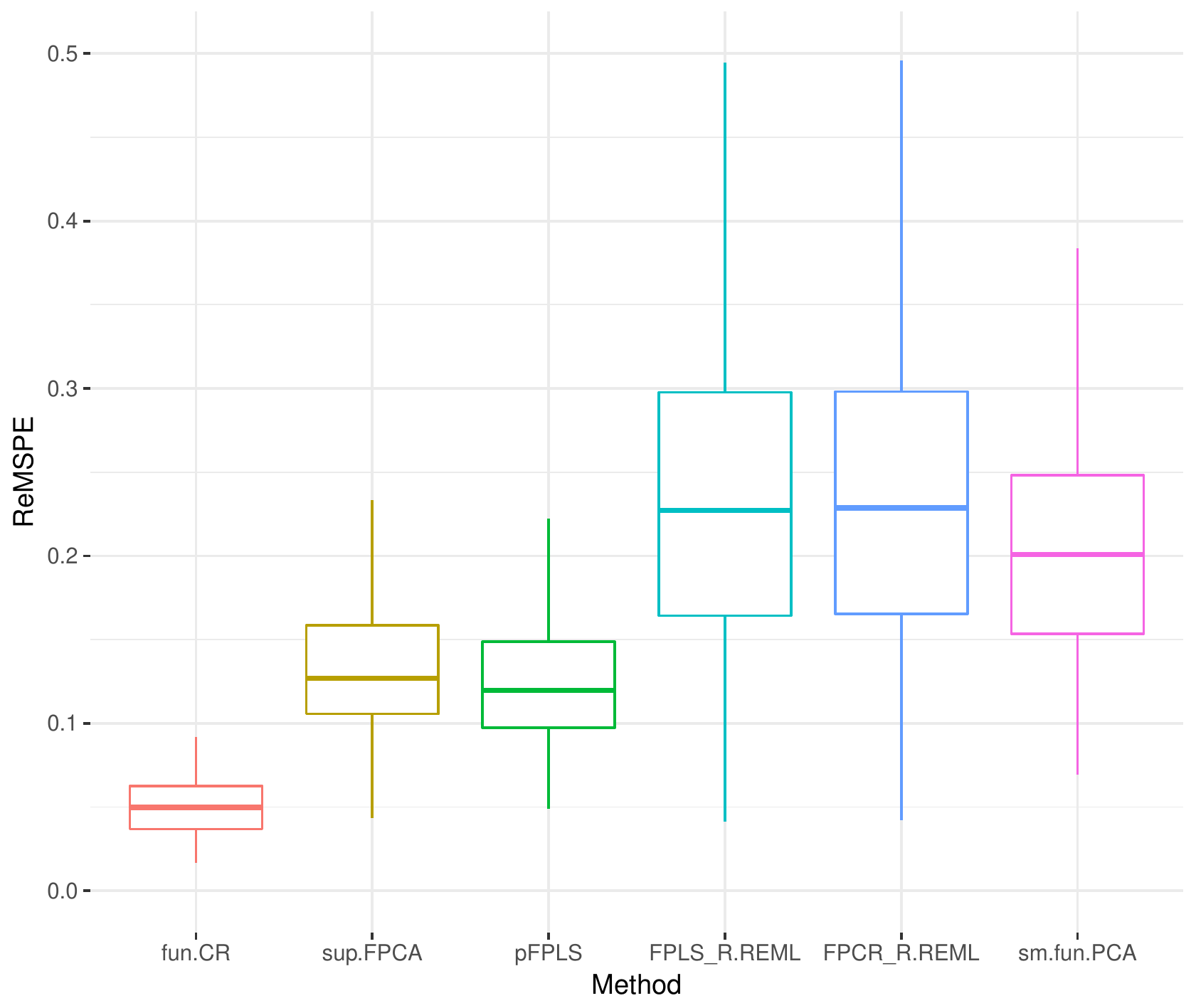}
		\caption{First 200 repeats}
	\end{subfigure}
	\caption{Boxplots of ReMSPEs of six methods for \textbf{Tecator\texttrademark~data (spectra vs. protein)}.
		In each subfigure,
		the six boxplots,
		from left to right,
		correspond to functional CR, supervised FPCA, pFPLS, FPLS$_R$-REML, FPCR$_R$-REML and smoothed functional PCA, respectively.
	} 
	\label{fig:boxplot.protein}
\end{figure}

We regress the fat, moisture and protein contents, respectively,
on the absorbance spectra.
In the first case (spectra vs. fat), 
the six approaches are roughly categorized into three groups 
in each subfigure of \autoref{fig:boxplot.fat}:
functional CR on the left end,
the three in the middle
(including supervised FPCA, pFPLS and FPLS$_R$-REML),
and another two on the very right
(i.e., FPCR$_R$-REML and smoothed functional PCA).
As shown in \autoref{fig:boxplot.fat},
supervised strategies are more favored than unsupervised ones.
In the cases of moisture (\autoref{fig:boxplot.moisture}) and protein (\autoref{fig:boxplot.protein}),
this phenomenon does not hold, 
but functional CR still takes the lead in terms of lower ReMSPEs,
even though we do not impose any penalty on the smoothness of functional continuum basis functions.


\section{Conclusion and discussion} \label{sec:conclusion}

Specially designed for scalar-on-function regression models,
the framework of functional CR encompasses the well-known functional PCR and functional PLS, etc..
We have given various equivalent forms of functional continuum basis functions
which lower the difficulty of optimization in the numerical implementation.
The consistency of estimators is demonstrated for the case of fixed $p$.
Verified in numerical studies and compared with five existing methods,
our strategy is overall competitive in terms of both estimation and prediction.

However,
our work is far from perfect.
The core of our algorithm is to locate the constrained global maximizer of $\ln Q_{p,\alpha}(\delta)$.
In \autoref{sec:numerical},
thanks to the simpleness of curves of $\ln Q_{p,\alpha}$,
we do not have to initiate the maximization with multiple start points.
Even so,
our implementation is still more involved than competitors when number of curves becomes larger; 
see \autoref{tab} for the time consumed for each method.
But it can always be worse;
it is possible for other datasets to be coupled with more complex curves for $\ln Q_{p,\alpha}$.  
In such cases,
we have to avoid the search being trapped in some local maxima.
We suggest using multiple initial values, 
a commonly adopted strategy.
But this significantly slow down the implementation of functional CR.
For instance,
under the same computing environment,
if we try 100 initial points in each maximization,
the seconds consumed by functional CR for the Tecator\texttrademark~data
would be over 30 times as many as the corresponding number posted in \autoref{tab}.

\begin{table}
	\centering\small
	\caption{
		Time consumed (seconds) by the analysis in \autoref{sec:numerical} with different approaches after 200 repeats
		(running on a laptop with 
		Intel\textsuperscript{\textregistered} Core\texttrademark\ i5-5200U CPU @ 2$\times$2.20 GHz and 8 GB RAM)
	}
	\label{tab}
	\begin{tabular}{c|rrr|rrr|r|r|r|r}
		\hline
		                        &\multicolumn{6}{c}{Simulation}
		                        &\multicolumn{1}{|c|}{Medfly}
		                        &\multicolumn{3}{c}{Tecator\texttrademark}	\\
		                        &\multicolumn{3}{c}{Scenario \ref{case:1st.eigen}}
		                        &\multicolumn{3}{c|}{Scenario \ref{case:3rd.eigen}}
  		                        &
		                        &\multicolumn{1}{c}{Fat}
		                        &\multicolumn{1}{c}{Moisture}
		                        &\multicolumn{1}{c}{Protein}  \\
		\hline
		SNR                     &20	&10	&2	&20	&10	&2	& 	&		&		&\\
		Number of curves		&35	&35	&35	&35	&35	&35	&50	&240	&240	&240\\						
		\hline
		functional CR           &94.7	&91.8	&95.1	&68.6	&69.4	&68.6	&176.5	&2303.4	&2393.9	&2104.8\\
		supervised FPCA         &91.0	&94.2	&93.8	&68.6	&68.9	&69.7	&55.8	&546.8	&576.8	&557.8\\
		pFPLS                   &653.2	&668.5	&679.8	&484.5	&492.3	&484.9	&394.0	&2090.3	&2283.6	&2029.5\\
		FPLS$_R$-REML           &66.4	&65.0	&64.7	&50.6	&49.6	&49.9	&34.7	&125.7	&110.3	&108.2\\
		FPCR$_R$-REML           &66.0	&63.7	&61.9	&51.3	&50.9	&55.0	&33.6	&120.4	&110.3	&111.3\\
		smoothed functional PCA &56.7 	&59.3 	&60.6 	&47.4 	&47.2 	&48.1	&34.1	&241.7	&207.9  &207.3\\        					
		\hline
	\end{tabular}
\end{table}

Last but not least,
functional CR possesses the potential to be further extended.
With a generalization analogous to that in \citet{BrooksStone1994},
it is hopeful to handle multiple responses simultaneously and even functional response.
Another possible direction of evolution is to enhance the robustness
by replacing variance and covariance terms with robust counterparts;
just like \citet{SerneelsFilzmoserCrouxEspen2005} did for CR.

\section*{Acknowledgment}

We are grateful to the editor, associate editor and anonymous reviewers
for their extremely valuable comments and constructive suggestions.
Also,
we would like to thank 
the Natural Sciences and Engineering Research Council of Canada (NSERC) for financial supports.

\appendix

\section[\ref{appendix:lemma}]{Lemmas}\label{appendix:lemma}

\autoref{lemma:existence} is the cornerstone of the existence of $w_{p,\alpha}$ and $\hat{w}_{p,\alpha}$.
\autoref{lemma:converge.T.star.hat},
essential in proving \autoref{thm:consistence},
reveals the convergence from the empirical objective function $\widehat{T}_{p,\alpha}^*$ to the theoretical one $T_{p,\alpha}^*$.
Their proofs are both left in \ref{appendix:proof}.

\begin{lemma}\label{lemma:existence}
	Suppose $C\subseteq L^2(\mathcal{T})$ is a bounded and weakly sequentially closed set.
	Suppose $f:C\rightarrow\mathbb{R}$ is weakly sequentially upper semi-continuous.  
	Then $f$ has a maximizer on $C$.
\end{lemma}

\begin{remark}
	Although the assumption of \autoref{lemma:existence} can be further relaxed,
	this less general version suffices for our needs in this paper;
	see Theorem 5.3 and Remark 5.4 
	in an unpublished 2013 technical report by A. Alexanderian
	(available at \url{https://aalexan3.math.ncsu.edu/articles/hilbert.pdf}).
\end{remark}

\begin{lemma}\label{lemma:converge.T.star.hat}
	Recall $T_{p,\alpha}^*(w)$ and $\widehat{T}_{p,\alpha}^*(w)$ both defined in \autoref{prop:equivalent.w.hat}.
	In case $\hat{w}_{j,\alpha}$ converges to $ w_{j,\alpha}$ in probability
	as $n$ goes to $\infty$ for $j=1,\ldots,p-1$,
	$\widehat{T}_{p,\alpha}^*(w)$ converges to $T_{p,\alpha}^*(w)$ in probability uniformly over the unit ball, i.e.,
	\[
	 	\lim_{n\rightarrow\infty}\Pr\left\{ \sup_{w:\|w\|\leq 1} \left| \widehat{T}_{p,\alpha}^*(w)-T_{p,\alpha}^*(w) \right|<\varepsilon\right\}=1,
		\quad
		\forall \varepsilon>0.
	\]
\end{lemma}

\section[\ref{appendix:proof}]{Proofs}\label{appendix:proof}

\begin{proof}[Proof of \autoref{lemma:existence}]
	Firstly prove that $f_0=\sup_{x\in C}f(x)<\infty$.
	To the contrary,
	suppose that $f_0=\infty$.
	Then there is a sequence $\{x_n\}\subset C$ such that $f(x_n)\geq n$ for each $n\in\mathbb{N}$.
	Deduced from the boundedness and weakly sequential closeness,
	the weakly sequential compactness of $C$ implies that
	$\{x_n\}$ must have a subsequence $\{x_{n_k}\}$
	which weakly converges to $x^*\in C$.
	Due to the weakly sequential upper semi-continuity of $f$,
	we have
	\[
		f(x^*)\geq\varlimsup_{k\rightarrow\infty}f(x_{n_k})\geq \varlimsup_{k\rightarrow\infty} n_k=\infty.
	\]
	This identity contradicts the range of $f$.
	
	Next, there always exists a sequence $\{x_n\}$ such that $\lim_{n\rightarrow\infty}f(x_n)=f_0$. 
	Find a weakly convergent sequence $\{x_{n_k}\}\subseteq \{x_n\}$ with limit $x^*\in C$.
	Thus,
	\[
		f_0=\sup_{x\in C}f(x)\geq f(x^*)\geq \varlimsup_{k\rightarrow\infty}f(x_{n_k})=\lim_{n\rightarrow\infty}f(x_n)=f_0.
	\]
	The sandwich rule indicates that $x^*\in C$ is a maximizer of $f$ on $C$ and completes this proof.
\end{proof}

\begin{proof}[Proof of \autoref{lemma:converge.T.star.hat}]
	The proof consists of three steps.
	First follow \citet[Eq. (5.1)]{DelaigleHall2012} to conclude that,
	as $n\rightarrow\infty$,
	\[
		\widehat{V}_{X^{(p,\alpha)}}(\beta)\overset{\text{P}}{\longrightarrow}V_{X^{(p,\alpha)}}(\beta) 
		\quad\text{and}\quad
		\hat{v}_{X^{(p,\alpha)}}\overset{\text{P}}{\longrightarrow}v_{X^{(p,\alpha)}}.
	\]
	Moreover,
	$\forall\varepsilon>0$,
	$\exists\delta>0$ such that
	\[
		\left\{\int_{\mathcal{T}}\int_{\mathcal{T}}
			\left( 
				\hat{v}_{X^{(p,\alpha)}}-\hat{v}_{\widehat{X}^{(p,\alpha)}}
			\right)^2>\varepsilon 
		\right\}
		\subseteq
		\left\{ 
			\int_{\mathcal{T}}\left( X^{(p,\alpha)}- \widehat{X}^{(p,\alpha)} \right)^2> \delta 
		\right\}
	\]
	and
	\begin{align*}
		\left\{\int_{\mathcal{T}}\left( 
			\widehat{V}_{\widehat{X}{(p,\alpha)}}(\beta)-\widehat{V}_{X^{(p,\alpha)}}(\beta) 
			\right)^2>\varepsilon \right\}
		&\subseteq
		\left\{ 
			\int_{\mathcal{T}}\left( X^{(p,\alpha)}- \widehat{X}^{(p,\alpha)} \right)^2
			> \delta \right\}.
	\end{align*}
	The continuous mapping theorem guarantees the convergence in probability of $\widehat{X}^{(p,\alpha)}(t)$ to $X^{(p,\alpha)}(t)$
	and further yields that
	\[
		\widehat{V}_{\widehat{X}^{(p,\alpha)}}(\beta)\overset{\text{P}}{\longrightarrow}V_{X^{(p,\alpha)}}(\beta) 
		\quad\text{and}\quad
		\hat{v}_{\widehat{X}^{(p,\alpha)}}\overset{\text{P}}{\longrightarrow}v_{X^{(p,\alpha)}}.
	\]
	
	Recall 
	$\widehat{V}_{\widehat{X}^{(p,\alpha)}}
	=\widehat{V}_{\widehat{X}^{(p,\alpha)}}(s,t)
	=n^{-1}\sum_{i=1}^n\widehat{X}_i^{(p,\alpha)}(s)\widehat{X}_i^{(p,\alpha)}(t)$
	and
	$\widehat{V}_{\widehat{X}^{(p,\alpha)}}(\beta)
	=n^{-1}\sum_{i=1}^n\widehat{X}_i^{(p,\alpha)}\widehat{Y}_i^{(p,\alpha)}$.
	For convenience,
	write
	\[
		f_{p,\alpha}=f_{p,\alpha}(w)=\int_{\mathcal{T}}wV_{X^{(p,\alpha)}}(\beta)
		\quad\text{and}\quad
		g_{p,\alpha}=g_{p,\alpha}(w)=\int_{\mathcal{T}}wV_{X^{(p,\alpha)}}(w)
	\]
	and their empirical conterparts
	\[
		\hat{f}_{p,\alpha}=\hat{f}_{p,\alpha}(w)=\int_{\mathcal{T}}w\widehat{V}_{\widehat{X}^{(p,\alpha)}}(\beta)
		\quad\text{and}\quad
		\hat{g}_{p,\alpha}=\hat{g}_{p,\alpha}(w)=\int_{\mathcal{T}}w\widehat{V}_{\widehat{X}^{(p,\alpha)}}(w).
	\]
	By the Cauchy-Schwarz inequality,
	as $n\rightarrow\infty$,
	\begin{align*}
		\sup_{w:\|w\|\leq 1} \left| f_{p,\alpha}(w)-\hat{f}_{p,\alpha}(w) \right| 
		&= \sup_{w:\|w\|\leq 1}
			\left| \int_{\mathcal{T}}w \left(V_{X^{(p,\alpha)}}(\beta)-\widehat{V}_{\widehat{X}^{(p,\alpha)}}(\beta)\right) \right| \\
		&\leq\left\|
			V_{X^{(p,\alpha)}}(\beta)-\widehat{V}_{\widehat{X}^{(p,\alpha)}}(\beta)
		\right\|
		\overset{\text{P}}{\longrightarrow}0,
	\end{align*}
	and
	\begin{align*}
		\sup_{w:\|w\|\leq 1} \left| g_{p,\alpha}(w)-\hat{g}_{p,\alpha}(w) \right| 
		&= \sup_{w:\|w\|\leq 1}
			\left| \int_{\mathcal{T}}w \left(V_{X^{(p,\alpha)}}(w)-\widehat{V}_{\widehat{X}^{(p,\alpha)}}(w)\right) \right| \\
		&\leq \left( \int_{\mathcal{T}}\int_{\mathcal{T}}\left( 
			v_{X^{(p,\alpha)}}-\hat{v}_{\widehat{X}^{(p,\alpha)}}
			\right)^2 \right)^{1/2} \overset{\text{P}}{\longrightarrow}0.
	\end{align*}
	
	Next we deduce a continuous mapping theorem specific for uniform convergence in probability.
	Suppose $m$ is a continuous $\mathbb{R}\times\mathbb{R}\rightarrow\mathbb{R}$ function.
	For arbitrary $\varepsilon>0$,
	there are $w_{n,\varepsilon}\in\{w:\|w\|\leq 1\}$ and $\delta>0$ such that
	\begin{align*}
		\left\{\sup_{w:\|w\|\leq 1} 
			\left| m\left(f_{p,\alpha}(w),g_{p,\alpha}(w)\right)
				-m\left(\hat{f}_{p,\alpha}(w),\hat{g}_{p,\alpha}(w)\right) \right|>\varepsilon \right\}
		&\subseteq
		\left\{ \left| m\left(f_{p,\alpha}(w_{n,\varepsilon}),g_{p,\alpha}(w_{n,\varepsilon})\right)
			-m\left(\hat{f}_{p,\alpha}(w_{n,\varepsilon}),\hat{g}_{p,\alpha}(w_{n,\varepsilon})\right) \right| >\varepsilon \right\} \\
		&\subseteq
		\left\{ \left| f_{p,\alpha}(w_{n,\varepsilon})- \hat{f}_{p,\alpha}(w_{n,\varepsilon}) \right|^2
			+\left| g_{p,\alpha}(w_{n,\varepsilon})- \hat{g}_{p,\alpha}(w_{n,\varepsilon}) \right|^2 > \frac{\delta^2}{2} \right\} \\
		&\subseteq
		\left\{ \left| f_{p,\alpha}(w_{n,\varepsilon})- \hat{f}_{p,\alpha}(w_{n,\varepsilon}) \right| > \frac{\delta}{2} \right\}
		\cup 
		\left\{ \left| g_{p,\alpha}(w_{n,\varepsilon})- \hat{g}_{p,\alpha}(w_{n,\varepsilon}) \right| > \frac{\delta}{2} \right\} \\
		&\subseteq
		\left\{\sup_{w:\|w\|\leq 1} \left| f_{p,\alpha}(w)- \hat{f}_{p,\alpha}(w) \right|> \frac{\delta}{2} \right\}
		\cup
		\left\{\sup_{w:\|w\|\leq 1} \left| g_{p,\alpha}(w)- \hat{g}_{p,\alpha}(w) \right|> \frac{\delta}{2} \right\},
	\end{align*}
	which further indicates that
	\[
		\lim_{n\rightarrow\infty}
		\Pr\left\{\sup_{w:\|w\|\leq 1} 
			\left| m\left(f_p(w),g_{p,\alpha}(w)\right)
				-m\left(\hat{f}_p(w),\hat{g}_{p,\alpha}(w)\right) \right|>\varepsilon \right\} =0.
	\]
	\autoref{lemma:converge.T.star.hat} follows the identities 
	$\widehat{T}_{p,\alpha}^*=\hat{f}_{p,\alpha}^2\cdot\hat{g}_{p,\alpha}^{\alpha/(\alpha-1)-1}$
	and $T_{p,\alpha}^*=f_{p,\alpha}^2\cdot g_{p,\alpha}^{\alpha/(\alpha-1)-1}$. 
\end{proof}

\begin{proof}[Proof of \autoref{prop:existence}]
	Denote the unit sphere and unit ball in $L^2(\mathcal{T})$ by
	\[
		S=\left\{w\in L^2(\mathcal{T}):\|w\|=1\right\}
		\quad\text{and}\quad
		B=\left\{w\in L^2(\mathcal{T}):\|w\|\leq 1\right\},
	\]
	respectively.
	Write
	\[
		W_{p-1,\alpha}^{\perp}
		=\left\{w\in L^2(\mathcal{T}): \int_{\mathcal{T}} wV_X(w_{j,\alpha})=0,\quad 1\leq j\leq p-1\right\}
	\]
	and 
	\[
		\widehat{W}_{p-1,\alpha}^{\perp}
		=\left\{w\in L^2(\mathcal{T}): \int_{\mathcal{T}} wV_X(\hat{w}_{j,\alpha})=0,\quad 1\leq j\leq p-1\right\}.
	\]
	Clearly,
	$W_{p-1,\alpha}^{\perp}\cap B$ is weakly sequentially closed and bounded
	and $T_{\alpha}(w)$ is weakly sequentially upper semi-continuous 
	if constrained on $W_{p-1,\alpha}^{\perp}\cap B$.
	According to \autoref{lemma:existence},
	$T_{\alpha}(w)$ has a maximizer within $W_{p-1,\alpha}^{\perp}\cap B$.
	This maximizer,
	say $w^*$, 
	must locate in
	$W_{p-1,\alpha}^{\perp}\cap S$,
	otherwise we can construct $w'=w^*/\sqrt{\int_{\mathcal{T}}w^{*2}}$
	with $T_{\alpha}(w')=\left(\int_{\mathcal{T}}w^{*2}\right)^{\alpha/(\alpha-1)}T_{\alpha}(w^*)>T_{\alpha}(w^*)$.
	Likewise,
	$\widehat{T}_{\alpha}(w)$ has a maximizer in 
	$\widehat{W}_{p-1,\alpha}^{\perp}\cap S$, too.
\end{proof}

\begin{proof}[Proof of \autoref{prop:expand}]
	Put aside two special cases:
	when $\alpha=0$,
	as stated in \autoref{sec:special},
	$\beta\propto w_{1,0}$;
	for $\alpha=1/2$,
	please synthesize (3.4) and (3.11) in \cite{DelaigleHall2012}.
	
	For $p\in\mathbb{N}$ and $\alpha\in(0,1/2)\cup(1/2,1)$,
	let 
	\begin{align*}
		f &=f(w)=\cov\left(Y, \int_{\mathcal{T}}Xw\right)=\int_{\mathcal{T}}wV_X(\beta),\\
		g &=g(w)=\int_{\mathcal{T}}wV_X(w),\\
		h &=h(w)=\int_{\mathcal{T}}w^2,\\
		e_j &=e_j(w)=2\int_{\mathcal{T}}wV_X(w_{j,\alpha}),\quad j=1,\ldots,p-1.
	\end{align*}
	Then $T_{\alpha}=f^2\cdot g^{\frac{\alpha}{1-\alpha}-1}$.
	The Lagrange multiplier rule for Banach spaces \cite[pp. 270--271]{Zeidler1995} ensures that
	there are real numbers $\delta_1,\ldots,\delta_p$,
	for each $w\in L^2(\mathcal{T})$,
	\begin{equation}\label{eq:lagrange}
		f(w_{p,\alpha})g^{\frac{\alpha}{1-\alpha}-2}(w_{p,\alpha})
		\left(2g(w_{p,\alpha})\DD f(w_{p,\alpha})(w)
			+\left(\frac{\alpha}{1-\alpha}-1\right)f(w_{p,\alpha})\DD g(w_{p,\alpha})(w)
		\right)
		=\delta_p\DD h(w_{p,\alpha})(w)
		+\sum_{j=1}^{p-1}\delta_j\DD e_j(w_{p,\alpha})(w),
	\end{equation}
	where $\DD f(w_{p,\alpha})$, $\DD g(w_{p,\alpha})$, $\DD h(w_{p,\alpha})$, and $\DD e_j(w_{p,\alpha})$,
		all surjections from $L^2(\mathcal{T})$ to $\mathbb{R}$,
		are the first-order (Fr\'{e}chet) derivatives of $f$, $g$, $h$, and $e_j$ 
		evaluated at $w_{p,\alpha}$, respectively;
		in particular,
		for $w\in L^2(\mathcal{T})$,
	\begin{align*}
		\DD f(w_{p,\alpha})(w)
			&=\int_{\mathcal{T}}wV_X(\beta), \\
		\DD g(w_{p,\alpha})(w)
			&=2\int_{\mathcal{T}}wV_X(w_{p,\alpha}), \\
		\DD h(w_{p,\alpha})(w)
			&=2\int_{\mathcal{T}}ww_{p,\alpha}, \\
		\DD e_j(w_{p,\alpha})(w)
			&=2\int_{\mathcal{T}}wV_X(w_{j,\alpha}),\quad j=1,\ldots,p-1.
	\end{align*}
	The arbitrariness of $w$ in Eq. \eqref{eq:lagrange} entails that
	\begin{equation}\label{eq:wp.Vx}
		f(w_{p,\alpha})g^{\frac{\alpha}{1-\alpha}-2}(w_{p,\alpha})
		\left(2g(w_{p,\alpha})V_X(\beta)
			+\left(\frac{\alpha}{1-\alpha}-1\right)f(w_{p,\alpha})V_X(w_{p,\alpha})
		\right)
		=\delta_p w_{p,\alpha}
		+\sum_{j=1}^{p-1}\delta_j V_X(w_{j,\alpha}).
	\end{equation}
	Cases of $\left(\frac{\alpha}{1-\alpha}-1\right)f^2(w_{p,\alpha})g^{\frac{\alpha}{1-\alpha}-1}(w_{p,\alpha})=0$
	and $\gamma_p=0$ are both eliminated:
	the former one corresponds to the uninteresting minimum of $T_{\alpha}$,
	while the latter one leads to the unconstrained maximizer of $T_{\alpha}$ 
	which actually never falls on the unit sphere.
	By Fredholm's theorems (see, e.g. \cite{Fredholm1903,Khvedelidze2011}),
	solve the integral equation \eqref{eq:wp.Vx}
	and acquire
	\[
		w_{p,\alpha}=U_{p,\alpha}\left(\gamma_p \beta+\sum_{j=1}^{p-1}\gamma_j w_{j,\alpha}\right),
	\]
	where $U_{p,\alpha}:L^2(\mathcal{T})\rightarrow L^2(\mathcal{T})$ 
	takes $w$ to $\left((V_X+\gamma_0 I)^{-1}\circ V_X\right)(w)$
	with $\gamma_0=\gamma_0(p,\alpha)\in\mathbb{R}$ and identity operator $I$
	and where $\gamma_1,\ldots,\gamma_p$ accommodate the $p$ side-conditions \eqref{eq:fcr.constraint}.
	It follows that
	\[
		\text{span}\left\{w_{1,\alpha},\ldots,w_{p,\alpha}\right\}
		=\text{span}\left\{K_{1,\alpha}(\beta),\ldots,K_{p,\alpha}(\beta)\right\},
	\]
	where $K_{p,\alpha}=U_{p,\alpha}\circ\cdots\circ U_{1,\alpha}$,
	because $w_{p,\alpha}$ is representable in terms of 
	$K_{1,\alpha}(\beta),\ldots,K_{p,\alpha}(\beta)$ for each $p$
	and vice versa.
	
	At last we verify that
	$\beta\in \overline{\text{span}\left\{K_{1,\alpha}(\beta),K_{2,\alpha}(\beta),\ldots\right\}}$.
	Introduce orthogonal projection operator $P_p$ that takes 
	$w\in L^2(\mathcal{T})$
	to
	$\sum_{j=1}^{p}\left(\int_{\mathcal{T}}w w_{j,\text{FPC}}\right)w_{j,\text{FPC}}$.
	Write 
	$\beta_{p,\text{FPC}}=P_p(\beta)$.
	Now,
	\[
		\left(
			\left(\frac{\lambda_1}{\lambda_1+\gamma_0(1,\alpha)}I-\left(P_p\circ U_{1,\alpha}\right)\right)
			\circ\cdots\circ
			\left(\frac{\lambda_p}{\lambda_p+\gamma_0(p,\alpha)}I-\left(P_p\circ U_{p,\alpha}\right)\right)
		\right)
		\left(\beta_{p,\text{FPC}}\right)=0
	\]
	in which $\lambda_j$ is the $j$-th eigenvalue of $V_X$,
	implying that 
	\[
		\beta_{p,\text{FPC}}
		\in\text{span}\left\{\left(P_p\circ K_{1,\alpha}\right)\left(\beta_{p,\text{FPC}}\right),
			\ldots,\left(P_p\circ K_{p,\alpha}\right)\left(\beta_{p,\text{FPC}}\right)\right\}.
	\]
	In view of
	$\left(P_p\circ K_{j,\alpha}\right)\left(\beta_{p,\text{FPC}}\right)=\left(P_p\circ K_{j,\alpha}\right)(\beta)$
	for $1\leq j\leq p$ and $p\in\mathbb{N}$,
	after taking limits in the $L^2$ sense as $p\rightarrow\infty$ on both sides of the following formula 
	\[
		\beta_{p,\text{FPC}}\in
		\left\{
			P_p(w): w\in\overline{\text{span}\left\{K_{1,\alpha}(\beta),K_{2,\alpha}(\beta),\ldots\right\}}
		\right\},
	\]
	we accomplish the proof.
\end{proof}

\begin{proof}[Proof of \autoref{prop:alpha->1}]
	For simplicity,
	we assume that $\lambda_1>\lambda_2>\cdots>0$ are eigenvalues of operator $V$,
	i.e., there is no tie among them.
	Then 
	\[
		\int_{\mathcal{T}}w_{p,\text{FPC}}V_X(w_{q,\text{FPC}})=
		\begin{cases}
			\lambda_p & \text{if } p=q\\
			0 & \text{if } p\neq q.
		\end{cases}
	\]
	The proposition can be proved by mathematical induction.
	
	For any $w$ ($\neq w_{1,\text{FPC}}$) on $S$ with $\int_{\mathcal{T}}wV_X(w)>0$,
	there exists $\alpha_0>2/3$ such that,
	for all $\alpha\in(\alpha_0,1)$,
	\[
		0<\left(\frac{\int_{\mathcal{T}}wV_X(w)}{\lambda_1}\right)^{\frac{\alpha}{1-\alpha}-1}
		<\frac{{\cov}^2(Y-\E Y,\int_{\mathcal{T}} Xw_{1,\text{FPC}})}{{\cov}^2(Y-\E Y,\int_{\mathcal{T}} Xw)},
	\]
	because $0<\int_{\mathcal{T}}wV_X(w)/\lambda_1<1$ and ${\cov}^2(Y-\E Y,\int_{\mathcal{T}} Xw_{1,\text{FPC}})>0$.
	It follows that $T_{\alpha}(w_{1,\text{FPC}})/T_{\alpha}(w)>1$ for all $\alpha\in(\alpha_0,1)$
	and hence $w_{1,\alpha}=w_{1,\text{FPC}}$ as $\alpha\rightarrow 1$.
	
	Suppose we have $w_{1,\alpha}=w_{1,\text{FPC}},\ldots,w_{p-1,\alpha}=w_{p-1,\text{FPC}}$,
	for certain $p\geq 2$.
	For $w$ ($\neq w_{p,\text{FPC}}$) satisfying constraints \eqref{eq:fcr.constraint} 
	and $\int_{\mathcal{T}}wV_X(w)>0$,
	along with sufficiently large $\alpha$,
	the inequalities
	\[
		0<\left(\frac{\int_{\mathcal{T}}wV_X(w)}{\lambda_p}\right)^{\frac{\alpha}{1-\alpha}-1}
		<\frac{{\cov}^2\left(Y,\int_{\mathcal{T}} Xw_{p,\text{FPC}}\right)}{{\cov}^2\left(Y,\int_{\mathcal{T}} Xw\right)}
	\]
	always hold.
	Thus, as $\alpha\rightarrow 1$,
	$w_{p,\text{FPC}}=\argmax_wT_{p,\alpha}(w)$ subject to \eqref{eq:fcr.constraint}
	and hence $w_{p,\alpha}=w_{p,\text{FPC}}$.
\end{proof}

\begin{proof}[Proof of \autoref{prop:equivalent.w}]
	Define $S$ and $W_{p-1,\alpha}^{\perp}$ as in the proof of \autoref{prop:existence}.
	Apparently,
	$T_{\alpha}(w)=T_{p,\alpha}^*(w)$
	for all $w\in W_{p-1,\alpha}^{\perp}$.
	That is,
	$w_{p,\alpha}$ is also the solution to 
	\begin{align}
		\notag&\underset{w}{\text{maximize}} & & T_{p,\alpha}^*(w) \\
		\notag&\text{subject to} & & \|w\|=1\quad\text{and}\quad \\
		\label{eq:orthogonal}& & & \int_{\mathcal{T}}wV_X(w_{j,\alpha})=0, \quad 1\leq j\leq p-1.
	\end{align}
	For any $w\in S$,
	construct $w^*\in S$ proportional to
	\[
		w-\sum_{j=1}^{p-1}
		\frac{\int_{\mathcal{T}} wV_X(w_{j,\alpha})}{\int_{\mathcal{T}}w_{j,\alpha}V_X(w_{j,\alpha})}
		w_{j,\alpha}.
	\]
	Due to 
	\[
		\int_{\mathcal{T}}\left(w-
			\sum_{j=1}^{p-1}\frac{\int_{\mathcal{T}} wV_X(w_{j,\alpha})}{\int_{\mathcal{T}}w_{j,\alpha}V_X(w_{j,\alpha})}w_{j,\alpha}\right)^2
		\leq 1
	\]
	and $\alpha/(\alpha-1)<0$ (excluding the trivial case $\alpha=0$),
	it is easy to verify that $w^* \in W_{p-1,\alpha}^{\perp}$
	and 
	\[
		T_{p,\alpha}^*(w^*)=
		\left(\int_{\mathcal{T}}\left(w-
			\sum_{j=1}^{p-1}
			\frac{\int_{\mathcal{T}} wV_X(w_{j,\alpha})}{\int_{\mathcal{T}}w_{j,\alpha}V_X(w_{j,\alpha})}
			w_{j,\alpha}\right)^2
		\right)^{\frac{\alpha}{\alpha-1}}
		T_{p,\alpha}^*(w)\geq T_{p,\alpha}^*(w).
	\]
	The inequality is an equality only when $w\in W_{p-1,\alpha}^{\perp}$,
	In other words,
	it suffices to drop side-conditions \eqref{eq:orthogonal}
	when maximizing $T_{p,\alpha}^*(w)$ subject to $\|w\|=1$.
\end{proof}

\begin{proof}[Proof of \autoref{prop:equivalent.w.hat}]
	Change every population values in the proof of \autoref{prop:equivalent.w} into empirical counterparts.
\end{proof}

\begin{proof}[Proof of \autoref{prop:explicit.w}]
	For $p\in\mathbb{N}$,
	let $h=h(w)=\int_{\mathcal{T}}w^2$, 	
	\[
		f_{p,\alpha}=f_{p,\alpha}(w)
		=\cov\left(Y^{(p,\alpha)}, \int_{\mathcal{T}}X^{(p,\alpha)}w\right)=\int_{\mathcal{T}}wV_{X^{(p,\alpha)}}(\beta),
	\]
	and
	\[
		g_{p,\alpha}=g_{p,\alpha}(w)
		=\int_{\mathcal{T}}wV_{X^{(p,\alpha)}}(w).
	\]
	Then $T_{p,\alpha}=f_{p,\alpha}^2\cdot g_{p,\alpha}^{\frac{\alpha}{1-\alpha}-1}$ and
	$w_{p,\alpha}$ defined as \eqref{eq:wp.fc} must be a solution to the constrained optimization problem
	\begin{align*}
		&\underset{w}{\text{maximize}} & & f_{p,\alpha}^2(w)\\
		&\text{subject to} & & 
			g_{p,\alpha}(w)=g_0
			\quad\text{and}\quad
			h(w)=1
	\end{align*}
	for certain $g_0\in(0,\lambda_1^{(p,\alpha)}]$,
	where $\lambda_j^{(p,\alpha)}$ is the $j$-th largest eigenvalue of operator $V_{X^{(p,\alpha)}}$
	with corresponding eigenfunction $\phi_j^{(p,\alpha)}$.
	
	Check the case with $g_0=\lambda_1^{(p,\alpha)}>0$
	(i.e., the functional principal component basis).
	Provided that $\lambda_1^{(p,\alpha)}$ has multiplicity $=m\geq 1$,
	we can write $w_{p,\alpha}=\sum_{j=1}^m a_j\phi_j^{(p,\alpha)}$,
	where $a_1,\ldots, a_m\in [-1,1]$ and $\sum_{j=1}^m a_j^2=1$.
	The Cauchy-Schwarz inequality implies that
	the maximum of
	\[
		f_{p,\alpha}^2(w)
		=\left(\sum_{j=1}^m a_j\int_{\mathcal{T}}\phi_j^{(p,\alpha)}V_{X^{(p,\alpha)}}(\beta)\right)^2
		=\left(\sum_{j=1}^m a_j\lambda_j^{(p,\alpha)}
			\int_{\mathcal{T}}\beta\phi_j^{(p,\alpha)}\right)^2
	\]
	is achieved if and only if
	\[
		(a_1,\ldots, a_m)\propto 
		\left(
			\lambda_1^{(p,\alpha)}\int_{\mathcal{T}}\beta\phi_1^{(p,\alpha)}, \ldots,
			\lambda_m^{(p,\alpha)}\int_{\mathcal{T}}\beta\phi_m^{(p,\alpha)}
		\right).
	\]
	Therefore,
	\[
		w_{p,\alpha}
		\propto
		\sum_{j=1}^{\infty}
			\frac{\lambda_j^{(p,\alpha)}\int_{\mathcal{T}}\beta\phi_j^{(p,\alpha)}}
			{\lambda_j^{(p,\alpha)}+\lambda_1^{(p,\alpha)}/\delta^{(p,\alpha)}}
			\phi_j^{(p,\alpha)}
		\quad\text{as }\delta^{(p,\alpha)}\rightarrow -1.
	\]
			
	Unless $g_0=\lambda_1^{(p,\alpha)}>0$,
	apply the Lagrange multiplier rule for Banach spaces as in the proof of \autoref{prop:expand}
	and arrive at
	\[
		f_{p,\alpha}(w_{p,\alpha})V_{X^{(p,\alpha)}}(\beta)
		=\delta_1 V_{X^{(p,\alpha)}}(w_{p,\alpha})+\delta_2 w_{p,\alpha},
	\]
	with $\delta_1, \delta_2\in\mathbb{R}$.
	$\delta_2$ must be nonzero as the solution to the unconstraint optimization problem $\max T_{p,\alpha}^*$ 
	never falls on the unit sphere.
	Also, 
	we rule out the case $f_{p,\alpha}(w_{p,\alpha})=0$ which corresponds to the uninteresting minimum of $T_{p,\alpha}^*$.
	
	If $\delta_1=0$,
	the functional continuum basis reduces to functional PLS basis and
	$w_{p,\alpha}\propto V_{X^{(p,\alpha)}}(\beta)$.
	When $\delta^{(p,\alpha)}$ is close enough to $0$,
	$\lambda_1^{(p,\alpha)}/\delta^{(p,\alpha)}$ becomes dominant over $\lambda_j^{(p,\alpha)}$ for all $j$,
	i.e.,
	$\lambda_j^{(p,\alpha)}+\lambda_1^{(p,\alpha)}/\delta^{(p,\alpha)}$
	and $\lambda_{j'}^{(p,\alpha)}+\lambda_1^{(p,\alpha)}/\delta^{(p,\alpha)}$ approach each other for all $j\neq j'$.
	Accordingly, 
	\begin{align*}
		w_{p,\alpha}
			&\propto
				\sum_{j=1}^{\infty}
					\lambda_j^{(p,\alpha)}\left(\int_{\mathcal{T}}\beta\phi_j^{(p,\alpha)}\right)\phi_j^{(p,\alpha)}\\
			&\propto 
				\sum_{j=1}^{\infty}
					\frac{\lambda_j^{(p,\alpha)}\left(\int_{\mathcal{T}}\beta\phi_j^{(p,\alpha)}\right)}
						{\lambda_j^{(p,\alpha)}+\lambda_1^{(p,\alpha)}/\delta^{(p,\alpha)}}
					\phi_j^{(p,\alpha)}
		\quad\text{as }\delta^{(p,\alpha)}\rightarrow 0.
	\end{align*}
	
	In the case with nonzero $\delta_1$,
	solving the following inhomogeneous Fredholm integral equation with respect to $w_{p,\alpha}$,
	\[
		\frac{f_{p,\alpha}(w_{p,\alpha})}{\delta_1}V_{X^{(p,\alpha)}}(\beta)
		=\frac{\delta_2}{\delta_1}w_{p,\alpha}+V_{X^{(p,\alpha)}}(w_{p,\alpha}),
	\]
	we also obtain the solution
	\[
		w_{p,\alpha}
			\propto 
				\sum_{j=1}^{\infty}
					\frac{\lambda_j^{(p,\alpha)}\left(\int_{\mathcal{T}}\beta\phi_j^{(p,\alpha)}\right)}
						{\lambda_j^{(p,\alpha)}+\lambda_1^{(p,\alpha)}/\delta^{(p,\alpha)}}
					\phi_j^{(p,\alpha)}
,
	\]
	where $\delta^{(p,\alpha)}=\delta_1\lambda_1^{(p,\alpha)}/\delta_2$.
	The existence and uniqueness of this solution is guaranteed by Fredholm's theorems
	which hold here because $v_{X^{(p,\alpha)}}\in L^2(\mathcal{T}\times\mathcal{T})$.
	
	The last phase of this proof is to ascertain that $\delta^{(p,\alpha)}\notin (-\infty, -1)$.
	Without loss of generality,
	assume that $\int_{\mathcal{T}}\phi_j^{(p,\alpha)}V_{X^{(p,\alpha)}}(\beta)\geq 0$ for all $j$,
	otherwise we can use $-\phi_j^{(p,\alpha)}$ instead.
	$\sum_{j=1}^{\infty}\lambda_j^{(p,\alpha)}<\infty$ is a property of Hilbert-Schmidt operator $v_{X^{(p,\alpha)}}$, 
	further indicating that,
	if $\delta^{(p,\alpha)}\in (-\infty, -1)$,
 	then there must exist $j_0$ such that 
 	$\lambda_{j_0}^{(p,\alpha)}+\lambda_1^{(p,\alpha)}/\delta^{(p,\alpha)}$ is negative.
 	Under this circumstance, 
 	changing the sign of it will increase $f_{p,\alpha}^2(w_{p,\alpha})$
 	without altering $g_{p,\alpha}(w_{p,\alpha})$ or violating the unit norm constraint.
 	This contradicts the definition of $w_{p,\alpha}$
 	and completes the proof.
\end{proof}

\begin{proof}[Proof of \autoref{thm:consistence}]
	We resort to an argument similar to the proof adopted by \citet[Theorem 4.1.1]{Amemiya1985}
	and extend it from the finite-dimensional setting to the functional context.
	The unit ball $B$ is as defined in the proof of \autoref{prop:existence}.
	Start with $p=1$ and let
	$N$ be a neighborhood in $L^2(\mathcal{T})$ containing $w_{p,\alpha}$, namely,
	\[
		N_{1,\delta}=\left\{ w\in L^2(\mathcal{T}): \left\|w-w_{1,\alpha}\right\| < \delta \right\}, 
		\quad 0<\delta <2.
	\]
	Verify that $B\setminus N_{1,\delta}$ is weakly sequentially closed and bounded
	and $T_{1,\alpha}^*(w)$ is weakly sequentially upper semi-continuous within $B\setminus N_{1,\delta}$.
	Then \autoref{lemma:existence} guarantees the existence of $\max_{w\in B\setminus N_{1,\delta}}T_{1,\alpha}^*(w)$.
	
	Write
	\[
		\varepsilon = T_{1,\alpha}^*(w_{p,\alpha})-\max_{w\in B\setminus N_{1,\delta}}T_{1,\alpha}^*(w)>0
	\]
	and observe that
	\begin{align*}
		\left\{ \sup_{w:\|w\|=1} \left| \widehat{T}_{1,\alpha}^*(w)-T_{1,\alpha}^*(w) \right|
			<\frac{\varepsilon}{2}  \right\}
		&\subseteq
		\left\{ T_{1,\alpha}^*(\hat{w}_{p,\alpha})>\widehat{T}_{1,\alpha}^*(\hat{w}_{1,\alpha})-\frac{\varepsilon}{2} \right\}
		\cup 
		\left\{ \widehat{T}_{1,\alpha}^*(w_{p,\alpha})>T_{1,\alpha}^*(w_{1,\alpha})-\frac{\varepsilon}{2} \right\} \\
		&\subseteq
		\left\{ T_{1,\alpha}^*(\hat{w}_{1,\alpha})>\widehat{T}_{1,\alpha}^*(w_{1,\alpha})-\frac{\varepsilon}{2} \right\}
		\cup 
		\left\{ \widehat{T}_{1,\alpha}^*(w_{1,\alpha})>T_{1,\alpha}^*(w_{1,\alpha})-\frac{\varepsilon}{2} \right\} \\
		&\subseteq
		\left\{ T_{1,\alpha}^*(\hat{w}_{1,\alpha})>T_{1,\alpha}^*(w_{1,\alpha})-\varepsilon \right\} \\
		&\subseteq
		\left\{ \hat{w}_{1,\alpha}\in N_{1,\delta} \right\}.
	\end{align*}
	Implied by \autoref{lemma:converge.T.star.hat},
	$\lim_{n\rightarrow \infty}\Pr\left\{ \hat{w}_{1,\alpha}\in N_{1,\delta} \right\}=1$.
	Considering the arbitrariness of $\delta$,
	we conclude that $\hat{w}_{1,\alpha}\overset{\text{P}}{\longrightarrow}w_{1,\alpha}$ as $n\rightarrow\infty$.
	In case the convergence of $\hat{w}_{1,\alpha},\ldots,\hat{w}_{p-1,\alpha}$ holds,
	the prerequisite of \autoref{lemma:converge.T.star.hat} is fulfilled.
	Mimicking the argument for $p=1$,
	we deduce $\hat{w}_{p,\alpha}\overset{\text{P}}{\longrightarrow}w_{p,\alpha}$ as $n\rightarrow\infty$.
	
	As for $\hat{\beta}_{p,\alpha}$ and $\hat{\eta}_{p,\alpha}(x)$,
	the convergence can be proved 
	after we recall their definitions in \eqref{eq:beta.hat.fc} and \eqref{eq:h.hat.fc}
	and employ the continuous mapping theorem for convergence in probability.
\end{proof}

\begin{proof}[Proof of \autoref{prop:explicit.w.hat}]
	Follow the identical argument in the proof for \autoref{prop:explicit.w}
	but substitute empirical items for the population counterparts.
	Meanwhile,
	take the following identity into consideration:
	\begin{align*}
		\hat{\lambda}_j^{(p,\alpha)}\int_{\mathcal{T}}\beta\hat{\phi}_j^{(p,\alpha)}
		&=\int_{\mathcal{T}}\beta\widehat{V}_{\widehat{X}^{(p,\alpha)}}\left(\hat{\phi}_j^{(p,\alpha)}\right)\\
		&=\widehat{\cov}\left(
				\int_{\mathcal{T}}\widehat{X}^{(p,\alpha)}\beta,
				\int_{\mathcal{T}}\widehat{X}^{(p,\alpha)}\hat{\phi}_j^{(p,\alpha)}
			\right)\\
		&=\widehat{\cov}\left(
				\widehat{Y}^{(p,\alpha)},
				\int_{\mathcal{T}}\widehat{X}^{(p,\alpha)}\hat{\phi}_j^{(p,\alpha)}
			\right).
	\end{align*}
\end{proof}



\bibliographystyle{abbrvnat} 
\bibliography{mybibfile}





\end{document}